\documentclass[a4,12pt]{article}
\usepackage{graphicx}
\begin{document}
\thispagestyle{empty}
%\vspace*{1cm}
\begin{center}
{\huge \bf {A Bird's-Eye View of Density-Functional Theory}}\\
\vspace{1cm}
Klaus Capelle\\
\vspace{0.5cm}
{\it Departamento de F\'{\i}sica e Inform\'atica\\
Instituto de F\'{\i}sica de S\~ao Carlos\\
Universidade de S\~ao Paulo\\
Caixa Postal 369, S\~ao Carlos, 13560-970 SP, Brazil\\}
%\vspace{1.5cm}
%{\small (fifth revision November 2006)}
\end{center}
\vspace{1.5cm}
{\em keywords:} density-functional theory, electronic-structure theory,
electron correlation, many-body theory, local-density approximation
\vspace{1.5cm}
%\vspace{6cm}
%\noindent
%Also available as {\it cond-mat/0211443}\\
%from {\it http://arXiv.org/archive/cond-mat}
%\title{A Bird's-Eye View of Density-Functional Theory}
%\author{Klaus Capelle\\
%Departamento de F\'{\i}sica e Inform\'atica\\
%Instituto de F\'{\i}sica de S\~ao Carlos\\
%Universidade de S\~ao Paulo\\
%Caixa Postal 369, S\~ao Carlos, 13560-970 SP, Brazil}
\begin{abstract}
This paper is the outgrowth of lectures the author gave at the Physics 
Institute and the Chemistry Institute of the University of S\~ao Paulo 
at S\~ao Carlos, Brazil, and at the VIII'th Summer School on Electronic 
Structure of the Brazilian Physical 
Society. It is an attempt to introduce density-functional theory (DFT) in a 
language accessible for students entering the field or researchers from other 
fields. It is not meant to be a scholarly review of DFT, but rather an informal
guide to its conceptual basis and some recent developments and advances. The 
Hohenberg-Kohn theorem and the Kohn-Sham equations are discussed in some 
detail. Approximate density functionals, selected aspects of applications 
of DFT, and a variety of extensions of standard DFT are also discussed, albeit 
in less detail. Throughout it is attempted to provide a balanced treatment of 
aspects that are relevant for chemistry and aspects relevant for physics, but 
with a strong bias towards conceptual foundations. The paper is intended to be 
read before (or in parallel with) one of the many excellent more technical 
reviews available in the literature. 
\end{abstract}

\newcommand{\be}{\begin{equation}}
\newcommand{\ee}{\end{equation}}
\newcommand{\bea}{\begin{eqnarray}}
\newcommand{\eea}{\end{eqnarray}}
\newcommand{\bi}{\bibitem}
\newcommand{\eps}{\epsilon}
\renewcommand{\r}{({\bf r})}
\newcommand{\rp}{({\bf r'})}
\newcommand{\rpp}{({\bf r''})}
\newcommand{\rrp}{({\bf r},{\bf r}')}
\newcommand{\ua}{\uparrow}
\newcommand{\da}{\downarrow}
\newcommand{\la}{\langle}
\newcommand{\ra}{\rangle}
\newcommand{\s}{\sigma}

%\maketitle

\newpage
\tableofcontents
\newpage

\section{Preface}
\label{preface}
\hspace{10mm}

This paper is the outgrowth of lectures the author gave at the Physics 
Institute and the Chemistry Institute of the University of S\~ao Paulo 
at S\~ao Carlos, Brazil, and at the VIII'th Summer School on Electronic 
Structure of the Brazilian Physical Society \cite{book}. 
The main text is a description of density-functional theory (DFT) at a level 
that should be accessible for students entering the field or researchers from 
other fields. A large number of footnotes provides additional comments and 
explanations, often at a slightly higher level than the main text. A 
reader not familiar with DFT is advised to skip most of the footnotes,
but a reader familiar with it may find some of them useful.

The paper is not meant to be a scholarly review of DFT, but rather an informal
guide to its conceptual basis and some recent developments and advances.
The Hohenberg-Kohn theorem and the Kohn-Sham equations are discussed in some
detail. Approximate density functionals, selected aspects of applications 
of DFT, and a variety of extensions of standard DFT are also discussed, albeit 
in less detail. Throughout it is attempted to provide a balanced treatment of 
aspects that are relevant for chemistry and aspects relevant for physics, but 
with a strong bias towards conceptual foundations. The text is intended to be 
read before (or in parallel with) one of the many excellent more technical 
reviews available in the literature. The author apologizes to all researchers
whose work has not received proper consideration. The limits of the author's 
knowledge, as well as the limits of the available space and the nature of the 
intended audience, have from the outset prohibited any attempt at 
comprehensiveness.\footnote{A first version of this text was published in
2002 as a chapter in the proceedings of the {\em VIII'th Summer School on 
Electronic Structure of the Brazilian Physical Society} \cite{book}. The text 
was unexpectedly well received, and repeated requests from users prompted the 
author to electronically publish revised, updated and extended versions in the
preprint archive {\em http://arxiv.org/archive/cond-mat}, where the second
(2003), third (2004) and fourth (2005) versions were deposited under the 
reference number cond-mat/0211443. The present fifth (2006) version of this 
text, published in the {\em Brazilian Journal of Physics}, is approximately
50\% longer than the first. Although during the consecutive revisions many
embarrassing mistakes have been removed, and unclear passages improved upon, 
many other doubtlessly remain, and much beautiful and important work has 
not been mentioned even in passing. The return from electronic publishing
to printed publishing, however, marks the completion of a cycle, and is
intended to also mark the end of the author's work on the {\em Bird's-Eye 
View of Density-Functional Theory}.}

\section{What is density-functional theory?}
\label{intro}
\hspace{10mm}

Density-functional theory is one of the most popular and successful 
quantum mechanical approaches to matter. It is nowadays routinely
applied for calculating, e.g., the binding energy of molecules in chemistry
and the band structure of solids in physics.
First applications relevant for fields traditionally considered more distant
from quantum mechanics, such as biology and mineralogy are beginning to appear.
Superconductivity, atoms in the focus of strong laser pulses, relativistic
effects in heavy elements and in atomic nuclei, classical liquids, and
magnetic properties of alloys have all been studied with DFT.

DFT owes this versatility to the generality of its fundamental concepts
and the flexibility one has in implementing them. In spite of this
flexibility and generality, DFT is based on quite a rigid conceptual 
framework. This section introduces some aspects of this framework in general 
terms. The following two sections, \ref{manybody} and \ref{singlebody}, then 
deal in detail with two core elements of DFT, the Hohenberg-Kohn theorem and 
the Kohn-Sham equations.  The final two sections, \ref{approx} and 
\ref{extensions}, contain a (necessarily less detailed) description of 
approximations typically made in practical DFT calculations, and of some 
extensions and generalizations of DFT.

To get a first idea of what density-functional theory is about, it is 
useful to take a step back and recall some elementary quantum mechanics. 
In quantum mechanics we learn that all information we can possibly have 
about a given system is contained in the system's wave function, $\Psi$.
Here we will exclusively be concerned with the electronic structure
of atoms, molecules and solids. The nuclear degrees of freedom (e.g.,
the crystal lattice in a solid) appear only in the form of a potential 
$v({\bf r})$ acting on the electrons, so that the wave function depends
only on the electronic coordinates.\footnote{This is the so-called
Born-Oppenheimer approximation. It is common to call $v\r$ a `potential' 
although it is, strictly speaking, a potential energy.}
Nonrelativistically, this wave function is calculated from Schr\"odinger's
equation, which for a single electron moving in a potential $v\r$ reads
\be
\left[ -\frac{\hbar^2\nabla^2}{2m} + v\r\right] \Psi\r
=\eps \Psi\r.
\label{sbse}
\ee
If there is more than one electron (i.e., one has a many-body problem) 
Schr\"odinger's equation becomes
\be
\left[\sum_i^N \left( -\frac{\hbar^2\nabla_i^2}{2m} + v({\bf r}_i) \right)
+\sum_{i<j} U({\bf r}_i,{\bf r}_j) \right] 
\Psi({\bf r}_1,{\bf r}_2 \ldots, {\bf r}_N)  
=E \Psi({\bf r}_1,{\bf r}_2 \ldots, {\bf r}_N),
\label{mbse}
\ee
where $N$ is the number of electrons and $U({\bf r}_i,{\bf r}_j)$ is the
electron-electron interaction. For a Coulomb system (the only type of
system we consider here) one has
\be
\hat{U}=\sum_{i<j} U({\bf r}_i,{\bf r}_j)=
\sum_{i<j} \frac{q^2}{|{\bf r}_i-{\bf r}_j|}.
\label{clbdef}
\ee
Note that this is the same operator for any system of particles interacting
via the Coulomb interaction, just as the kinetic energy operator 
\be
\hat{T} = -\frac{\hbar^2}{2m} \sum_i \nabla^2_i
\ee
is the same for any nonrelativistic system.\footnote{For materials containing 
atoms with large atomic number $Z$, accelerating the electrons
to relativistic velocities, one must include relativistic effects by solving
Dirac's equation or an approximation to it. In this case the kinetic energy
operator takes a different form.} Whether our system is an atom,
a molecule, or a solid thus depends only on the potential $v({\bf r}_i)$.
For an atom, e.g.,
\be
\hat{V}= \sum_i v({\bf r}_i)= \sum_i \frac{Qq}{|{\bf r}_i-{\bf R}|},
\ee 
where $Q$ is the nuclear charge\footnote{In terms of the elementary charge
$e>0$ and the atomic number $Z$, the nuclear charge is $Q=Ze$ and the charge
on the electron is $q=-e$.} and ${\bf R}$ the nuclear position.
When dealing with a single atom, ${\bf R}$ is usually taken to be the zero
of the coordinate system. For a molecule or a solid one has
\be
\hat{V}=\sum_i v({\bf r}_i)= \sum_{ik}\frac{Q_kq}{|{\bf r}_i-{\bf R}_k|},
\label{expotl}
\ee 
where the sum on $k$ extends over all nuclei in the system, each with charge 
$Q_k=Z_k e$ and position ${\bf R}_k$. It is only the spatial arrangement
of the ${\bf R}_k$ (together with the corresponding boundary conditions)
that distinguishes, fundamentally, a molecule from a solid.\footnote{One 
sometimes says that $\hat{T}$ and $\hat{U}$ are `universal', while $\hat{V}$ is
system-dependent, or `nonuniversal'. We will come back to this terminology.}
Similarly, it is only through the term $\hat{U}$ that the (essentially simple)
single-body quantum mechanics of Eq.~(\ref{sbse}) differs from the extremely 
complex many-body problem posed by Eq.~(\ref{mbse}).
These properties are built into DFT in a very fundamental way.

The usual quantum-mechanical approach to Schr\"odinger's equation (SE) can be
summarized by the following sequence
\be
\fbox{$
v\r \stackrel{SE}{\Longrightarrow} \Psi({\bf r}_1,{\bf r}_2 \ldots, {\bf r}_N) 
\stackrel{\la \Psi| \ldots|\Psi\ra}{\Longrightarrow}
{\rm observables},
$}
\ee
i.e., one specifies the system by choosing $v\r$, plugs it into Schr\"odinger's
equation, solves that equation for the wave function $\Psi$, and then 
calculates observables by taking expectation values of operators with this 
wave function. One among the observables that are calculated in this way is 
the particle
density \be
n\r = N \int d^3r_2 \int d^3r_3 \ldots \int d^3r_N
\Psi^*({\bf r},{\bf r}_2 \ldots, {\bf r}_N)
\Psi({\bf r},{\bf r}_2 \ldots, {\bf r}_N).
\label{densdef}
\ee
Many powerful methods for solving Schr\"odinger's equation have been developed 
during decades of struggling with the many-body problem. In physics, for 
example, one has diagrammatic perturbation theory (based on Feynman diagrams 
and Green's functions), while in chemistry one often uses configuration
interaction (CI) methods, which are based on systematic expansion in Slater
determinants. A host of more special techniques also exists. The problem with 
these methods is the great demand they place on one's computational resources: 
it is simply impossible to apply them efficiently to large and complex systems.
Nobody has ever calculated the chemical properties of a 100-atom molecule with 
full CI, or the electronic structure of a real semiconductor using nothing but 
Green's functions.\footnote{A simple estimate of the computational complexity 
of this task is to imagine a real-space representation of $\Psi$ on a mesh, in 
which each coordinate is discretized by using 20 mesh points (which is not very
much). For N electrons, $\Psi$ becomes a function of $3N$ coordinates (ignoring 
spin, and taking $\Psi$ to be real), and $20^{3N}$ values are required to 
describe $\Psi$ on the mesh. The density $n\r$ is a function of three 
coordinates, and requires $20^{3}$ values on the same mesh. CI and the 
Kohn-Sham formulation of DFT additionally employ sets of single-particle 
orbitals. $N$ such orbitals, used to build the density, require $20^{3} N$ 
values on the same mesh. (A CI calculation employs also unoccupied orbitals, 
and requires more values.) For $N=10$ electrons, the many-body wave function 
thus requires $20^{30}/20^3 \approx 10^{35}$ times more storage space than 
the density, and $20^{30}/(10 \times 20^3) \approx 10^{34}$ times more than
sets of single-particle orbitals. Clever use of symmetries can reduce these 
ratios, but the full many-body wave function remains unaccessible for real 
systems with more than a few electrons.} 

It is here where DFT provides a viable alternative, less accurate 
perhaps,\footnote{Accuracy is a relative term. As a theory, DFT is formally 
exact. Its performance in actual applications depends on the quality of the
approximate density functionals employed. For small numbers of particles,
or systems with special symmetries, essentially exact solutions of
Schr\"odinger's equation can be obtained, and no approximate functional
can compete with exact solutions. For more realistic systems, modern (2005)
sophisticated density functionals attain rather high accuracy. Data on atoms
are collected in  Table~\ref{table1} in Sec.~\ref{semilocal}. Bond-lengths
of molecules can be predicted with an average error of less than $0.001$nm, 
lattice constants of solids with an average error of less than $0.005$nm, 
and molecular energies to within less than $0.2$eV \cite{metaggatests}.
(For comparison: already a small molecule, such as water, has a total 
energy of $2081.1$eV). On the other hand, energy gaps in solids can 
be wrong by 100\%!} but 
much more versatile. DFT explicitly recognizes that nonrelativistic Coulomb 
systems differ only by their potential $v\r$, and supplies a prescription for 
dealing with the universal operators $\hat{T}$ and $\hat{U}$ 
once and for all.\footnote{We will see that in practice this prescription 
can be implemented only approximately. Still, these approximations retain a
high degree of universality in the sense that they often work well for more
than one type of system.} 
Furthermore, DFT provides a way to systematically map the many-body problem, 
with $\hat{U}$, onto a single-body problem, without $\hat{U}$. All this is done
by promoting the particle density $n\r$ from just one among many observables 
to the status of key variable, on which the calculation of all other 
observables can be based. This approach forms the basis of the large majority
of electronic-structure calculations in physics and chemistry. Much of what
we know about the electrical, magnetic, and structural properties of
materials has been calculated using DFT, and the extent to which DFT has
contributed to the science of molecules is reflected by the 1998 Nobel Prize
in Chemistry, which was awarded to Walter Kohn \cite{rmp1}, the founding 
father of DFT, and John Pople \cite{rmp2}, who was instrumental in 
implementing DFT in computational chemistry.

The density-functional approach can be summarized by the sequence
\be
\fbox{$
n\r \Longrightarrow \Psi({\bf r}_1,\ldots,{\bf r}_N) \Longrightarrow v\r,
$}
\ee
i.e., knowledge of $n\r$ implies knowledge of the wave function and the
potential, and hence of all other observables. 
Although this sequence describes the conceptual structure of DFT, it does
not really represent what is done in actual applications of it, which
typically proceed along rather different lines, and do not make explicit use
of many-body wave functions. The following chapters
attempt to explain both the conceptual structure and some of the many possible 
shapes and disguises under which this structure appears in applications.

The literature on DFT is large, and rich in excellent reviews and overviews.
Some representative examples of full reviews and systematic collections of
research papers are Refs.~[5-19].
The present overview of DFT is much less detailed and advanced than these 
treatments. Introductions to DFT that are more similar in spirit to the
present one (but differ in emphasis and selection of topics) are the 
contribution of Levy in Ref.~\cite{seminario}, the one of Kurth and Perdew 
in Refs.~\cite{joulbert} and \cite{primer}, 
and Ref.~\cite{makov} by Makov and Argaman.
My aim in the present text is to give a bird's-eye view of DFT in a language 
that should be accessible to an advanced undergraduate student who has 
completed a first course in quantum mechanics, in either chemistry or
physics. Many interesting details, proofs of theorems, illustrative 
applications, and exciting developments had to be left out, just as any
discussion of issues that are specific to only certain subfields of either
physics or chemistry. All of this, and much more, can be found in the
references cited above, to which the present little text may perhaps
serve as a prelude.

\section{DFT as a many-body theory}
\label{manybody}
\hspace{10mm}

\subsection{Functionals and their derivatives}
\label{functionals}

Before we discuss density-functional theory more carefully, let us introduce a
useful mathematical tool. Since according to the above sequence the wave 
function is determined by the density, we can write it as
$\Psi=\Psi[n]({\bf r}_1,{\bf r}_2,\ldots {\bf r}_N)$,
which indicates that $\Psi$ is a function of its $N$ spatial variables,
but a {\it functional} of $n\r$. 

{\em Functionals.} More generally, a functional $F[n]$ can be
defined (in an admittedly mathematically sloppy way) as a rule for going from
a function to a number, just as a function $y=f(x)$ is a rule ($f$) for going
from a number ($x$) to a number ($y$).
A simple example of a functional is the particle number,
\be
N = \int d^3r\, n\r = N[n],
\label{Ndef}
\ee
which is a rule for obtaining the number $N$, given the function $n\r$.
Note that the name given to the argument of $n$ is completely irrelevant,
since the functional depends on the {\it function} itself, not on its variable.
Hence we do not need to distinguish $F[n\r]$ from, e.g., $F[n\rp]$.
Another important case is that in which the functional depends on a parameter,
such as in
\be
v_H[n]\r = q^2 \int d^3r'\, \frac{n\rp}{|{\bf r}-{\bf r}'|},
\label{hartreepotl}
\ee
which is a rule that for any value of the parameter ${\bf r}$ associates a
value $v_H[n]\r$ with the function $n\rp$. This term is the so-called
Hartree potential, which we will repeatedly encounter below.

{\em Functional variation.}
Given a function of one variable, $y=f(x)$, one can think of two types of 
variations of $y$, one associated with $x$, the other with $f$. For a fixed 
functional dependence $f(x)$, the ordinary differential $dy$ measures how $y$ 
changes as a result of a variation $x\to x+dx$ of the variable $x$. This
is the variation studied in ordinary calculus.
Similarly, for a fixed point $x$, the functional variation 
$\delta y$ measures how the value $y$ at this point changes as a result of a
variation in the functional form $f(x)$. This is the variation studied in 
variational calculus. 

{\em Functional derivative.}
The derivative formed in terms of the ordinary differential, $df/dx$, measures
the first-order change of $y=f(x)$ upon changes of $x$, i.e., the slope of the 
function $f(x)$ at $x$:
\be
f(x+dx)=f(x)+ \frac{df}{dx} dx + O(dx^2).
\ee 
The functional derivative measures,
similarly, the first-order change in a functional upon a functional
variation of its argument:
\be
F[f(x)+\delta f(x)]=F[f(x)] + \int s(x)\, \delta f(x)\, dx + O(\delta f^2),
\ee
where the integral arises because the variation in the functional $F$ is 
determined by variations in the function at all points in space.
The first-order coefficient (or `functional slope') $s(x)$ is defined to be 
the functional derivative $\delta F[f]/\delta f(x)$.

The functional derivative allows us to study how a functional changes upon 
changes in the form of the function it depends on. Detailed rules for 
calculating functional derivatives are described in Appendix A of 
Ref.~\cite{parryang}. A general expression  for 
obtaining functional derivatives with respect to $n(x)$ of a functional
$F[n]=\int f(n,n',n'',n''',...;x) dx$, where primes indicate ordinary
derivatives of $n(x)$ with respect to $x$, is \cite{parryang}
\be
\frac{\delta F[n]}{\delta n(x)} = \frac{\partial f}{\partial n}
-\frac{d}{dx}\frac{\partial f}{\partial n'}
+\frac{d^2}{dx^2}\frac{\partial f}{\partial n''} 
-\frac{d^3}{dx^3}\frac{\partial f}{\partial n'''}
+...
\label{funcderiv}
\ee
This expression is frequently used in DFT to obtain $xc$ potentials from $xc$
energies.\footnote{ The use of functionals and their derivatives is not limited
to density-functional theory, or even to quantum mechanics. In classical 
mechanics, e.g., one expresses the Lagrangian ${\cal L}$ in terms of of 
generalized coordinates $q(x,t)$ and their temporal derivatives $\dot{q}(x,t)$,
and obtains the equations of motion from extremizing the action functional 
${\cal A}[q] = \int {\cal L}(q,\dot{q};t)dt$. The resulting equations of 
motion are the well-known Euler-Lagrange equations
 $0=\frac{\delta {\cal A}[q]}{\delta q(t)}=
\frac{\partial {\cal L}}{\partial q} 
-\frac{d}{dt}\frac{\partial {\cal L}}{\partial \dot{q}}$, 
which are a special case of Eq.~(\ref{funcderiv}).}

\subsection{The Hohenberg-Kohn theorem}
\label{hktheorem}

At the heart of DFT is the Hohenberg-Kohn (HK) theorem. This theorem states
that for ground states Eq.~(\ref{densdef}) can be inverted: given a 
{\it ground-state} density $n_0\r$ it is possible, in principle, to 
calculate the corresponding {\it ground-state} wave function 
$\Psi_0({\bf r}_1,{\bf r}_2 \ldots, {\bf r}_N)$. This means that $\Psi_0$ 
is a functional of $n_0$. Consequently, all ground-state observables
are functionals of $n_0$, too.
If $\Psi_0$ can be calculated from $n_0$ and vice versa, both
functions are equivalent and contain exactly the same information. 
At first sight this seems impossible: how can a function of one (vectorial)
variable ${\bf r}$ be equivalent to a function of $N$ (vectorial) variables
${\bf r}_1 \ldots {\bf r}_N$? How can one arbitrary variable contain the 
same information as $N$ arbitrary variables?

The crucial fact which makes this possible is that knowledge of $n_0\r$ implies
implicit knowledge of much more than that of an arbitrary function $f\r$.
The ground-state wave function $\Psi_0$ must not only reproduce the 
ground-state density, but also minimize the energy. 
For a given ground-state density $n_0\r$, we can write this requirement as
\be
E_{v,0}= \min_{\Psi\to n_0} \la \Psi |\hat{T} +\hat{U} +\hat{V}|\Psi \ra,
\label{csproof1}
\ee
where $E_{v,0}$ denotes the ground-state energy in potential $v\r$.
The preceding equation tells us that for a given density $n_0\r$ the 
ground-state wave function $\Psi_0$ is that which reproduces this $n_0\r$ 
{\em and} minimizes the energy. 

For an arbitrary density $n\r$, we define the functional
\be
E_v[n]=\min_{\Psi\to n}\la \Psi | \hat{T} + \hat{U} +\hat{V} | \Psi \ra.
\label{csproof2}
\ee
If $n$ is a density different from the ground-state density $n_0$ in potential
$v\r$, then the $\Psi$ that produce this $n$ are different from the 
ground-state wave function $\Psi_0$, and according to the variational 
principle the minimum obtained from $E_v[n]$ is higher than (or equal to) 
the ground-state energy $E_{v,0}=E_v[n_0]$. Thus, the functional $E_v[n]$ 
is minimized by the ground-state density $n_0$, and its value at the minimum 
is $E_{v,0}$.

The total-energy functional can be written as
\be
\fbox{$
E_v[n]= \min_{\Psi\to n}\la \Psi | \hat{T} + \hat{U} | \Psi \ra
+ \int d^3 r\, n\r v\r
=: F[n] + V[n],
$}
\label{csproof3}
\ee
where the internal-energy functional
$F[n]= \min_{\Psi\to n}\la \Psi | \hat{T} + \hat{U} | \Psi \ra$
is independent of the potential $v\r$, and thus determined only by the
structure of the operators $\hat{U}$ and $\hat{T}$. This universality of
the internal-energy functional allows us to define the ground-state wave 
function $\Psi_0$ as that antisymmetric $N$-particle function that delivers 
the minimum of $F[n]$ {\em and} reproduces $n_0$. If the ground state is 
nondegenerate (for the case of degeneracy see footnote 12), this double
requirement uniquely determines $\Psi_0$ in terms of $n_0\r$, without having
to specify $v\r$ explicitly.\footnote{Note that this is exactly the opposite
of the conventional prescription to specify the Hamiltonian via $v\r$, and
obtain $\Psi_0$ from solving Schr\" odinger's equation, without having
to specify $n\r$ explicitly.}

Equations (\ref{csproof1}) to (\ref{csproof3}) constitute the 
constrained-search proof of
the Hohenberg-Kohn theorem, given independently by M.~Levy \cite{levy} and 
E.~Lieb \cite{lieb}. The original proof by Hohenberg and Kohn \cite{hk}
proceeded by assuming that $\Psi_0$ was not determined uniquely by 
$n_0$ and showed that this produced a contradiction to the variational
principle. Both proofs, by constrained search and by contradiction, are 
elegant and simple. In fact, it is a bit surprising that it took 38 years from 
Schr\"odinger's first papers on quantum mechanics \cite{schroedinger} to 
Hohenberg and Kohn's 1964 paper containing their famous theorem \cite{hk}. 

Since 1964, the HK theorem has been thoroughly scrutinized, and several
alternative proofs have been found. One of these is the so-called `strong 
form of the Hohenberg-Kohn theorem', based on the inequality 
\cite{theophilou,stronghk,strange}
\be
\int d^3 r \Delta n\r \Delta v\r < 0.
\label{ineq}
\ee
Here $\Delta v\r$ is a change in the potential, and $\Delta n\r$ is the
resulting change in the density. We see immediately that if 
$\Delta v \neq 0$ we cannot have $\Delta n\r\equiv 0$, i.e., a change in
the potential must also change the density. This observation implies again
the HK theorem for a single density variable: there cannot be two local 
potentials with the same
ground-state charge density. A given $N$-particle ground-state density thus
determines uniquely the corresponding potential, and hence also the wave 
function. Moreover, (\ref{ineq})
establishes a relation between the signs of  $\Delta n\r$ and $\Delta v\r$:
if $\Delta v$ is mostly positive, $\Delta n\r$ must be mostly negative, so
that their integral over all space is negative. This additional information
is not immediately available from the two classic proofs, and is the
reason why this is called the `strong' form of the HK theorem.
Equation (\ref{ineq}) can be obtained along the lines of the standard HK
proof \cite{theophilou,stronghk}, but it can be turned into an independent 
proof of the HK theorem because it can also be derived perturbatively (see,
e.g., section 10.10 of Ref.~\cite{strange}).

Another alternative argument is valid only for Coulomb potentials. 
It is based on Kato's theorem,
which states \cite{kato,march2} that for such potentials the electron density 
has a cusp at the position of the nuclei, where it satisfies
\be
Z_k = \left. 
-\frac{a_0}{2 n\r}\frac{d n\r}{d r} \right|_{{\bf r}\to {\bf R}_k}.
\label{kato}
\ee
Here ${\bf R}_k$ denotes the positions of the nuclei, $Z_k$ their atomic
number, and $a_0=\hbar^2/me^2$ is the Bohr radius.
For a Coulomb system one can thus, in principle, read off all information 
necessary for completely specifying the Hamiltonian directly from examining
the density distribution: the integral over $n\r$ yields $N$, the total 
particle number; the position of the cusps of $n\r$ are the positions of 
the nuclei, ${\bf R}_k$; and the derivative of $n\r$ at these positions 
yields $Z_k$ by means of Eq.~(\ref{kato}). This is all one needs to specify 
the complete Hamiltonian of Eq.~(\ref{mbse}) (and thus implicitly all its 
eigenstates). In practice one almost never knows the density distribution 
sufficiently well to implement the search for the cusps and calculate the 
local derivatives. Still, Kato's theorem provides a vivid illustration of 
how the density can indeed contain sufficient information to completely 
specify a nontrivial Hamiltonian.\footnote{Note that, unlike the full
Hohenberg-Kohn theorem, Kato's theorem does apply only to superpositions of 
Coulomb potentials, and can therefore not be applied directly to the effective 
Kohn-Sham potential.}

For future reference we now provide a commented summary of the content of
the HK theorem. This summary consists of four statements:

{\bf (1)} The nondegenerate ground-state (GS) wave function is a unique 
functional of the GS density:\footnote{If the ground state is degenerate, 
several of the degenerate ground-state wave functions may produce the same 
density, so that a unique functional $\Psi[n]$ does not exist, but by 
definition these wave functions all yield the same energy, so that the 
functional $E_v[n]$ continues to exist and to be minimized by $n_0$. 
A universal functional $F[n]$ can also still be defined \cite{dftbook}.}
\be
\fbox{$
\Psi_0({\bf r}_1,{\bf r}_2 \ldots, {\bf r}_N)= \Psi[n_0\r]$.}
\ee
This is the essence of the HK theorem. As a consequence, the GS expectation
value of any observable $\hat{O}$ is a functional of $n_0\r$, too:
\be
\fbox{$
O_0 = O[n_0] = \la \Psi[n_0] | \hat{O} | \Psi[n_0] \ra$.}
\label{expecvalue}
\ee

{\bf (2)} Perhaps the most important observable is the GS energy. This 
energy
\be
E_{v,0} = E_v[n_0] = \la \Psi[n_0] | \hat{H} | \Psi[n_0] \ra,
\label{etot}
\ee
where $\hat{H}=\hat{T}+\hat{U}+\hat{V}$, has the variational property\footnote{The minimum of $E[n]$ is thus attained for the ground-state density. All 
other extrema of this functional correspond to densities of excited states,
but the excited states obtained in this way do not necessarily cover the 
entire spectrum of the many-body Hamiltonian \cite{perdewlevy}.}
\be
\fbox{$
E_v[n_0] \leq  E_v[n']$,}
\label{variational}
\ee
where $n_0$ is GS density in potential $\hat{V}$ and $n'$ is some other density.
This is very similar to the usual variational principle for wave functions.
From a calculation of the expectation value of a Hamiltonian with a trial wave 
function $\Psi'$ that is not its GS wave function $\Psi_0$ one can never obtain 
an energy below the true GS energy,
\be
E_{v,0}=E_{v}[\Psi_0] = \la \Psi_0 | \hat{H} | \Psi_0 \ra
\leq \la \Psi' | \hat{H} | \Psi' \ra = E_{v}[\Psi'].
\ee
Similarly, in exact DFT, if $E[n]$ for fixed $v_{ext}$ is evaluated for a 
density that is not the GS density of the system in potential $v_{ext}$, one 
never finds a result below the 
true GS energy. This is what Eq.~(\ref{variational}) says, and it is so 
important for practical applications of DFT that it is sometimes called
the {\it second Hohenberg-Kohn theorem} (Eq.~(\ref{expecvalue}) is the first 
one, then). 

In performing the minimization of $E_v[n]$ the constraint that the total 
particle number $N$ is an integer is taken into account by means of a
Lagrange multiplier, replacing the constrained minimization of $E_v[n]$ by
an unconstrained one of $E_v[n]-\mu N$. Since $N=\int d^3r n\r$, this leads to
\be
\frac{\delta E_v[n]}{\delta n\r} = \mu = \frac{\partial E}{\partial N},
\ee
where $\mu$ is the chemical potential. 

{\bf (3)} Recalling that the kinetic and interaction energies of a
nonrelativistic Coulomb system are described by universal operators, we can
also write $E_v$ as
\be
\fbox{$
E_v[n] = T[n] + U[n] + V[n] =F[n] + V[n]$,}
\ee
where $T[n]$ and $U[n]$ are {\it universal functionals} [defined as
expectation values of the type (\ref{expecvalue}) of $\hat{T}$ and 
$\hat{U}$], independent of $v\r$. 
On the other hand, the potential energy in a given potential $v\r$ 
is the expectation value of Eq.~(\ref{expotl}),
\be
V[n] = \int d^3r\, n\r v\r,
\label{vdef}
\ee
and obviously nonuniversal (it depends on $v\r$, i.e., on the system under 
study), but very simple: once the system is specified, i.e., $v\r$
is known, the functional $V[n]$ is known explicitly.

{\bf (4)} There is a fourth substatement to the HK theorem, which shows that 
if $v\r$ is not hold fixed, the functional $V[n]$ becomes universal:
the GS density determines not only the GS wave function $\Psi_0$,
but, up to an additive constant, also the potential $V=V[n_0]$.
This is simply proven by writing Schr\"odinger's equation as
\be
\hat{V} = \sum_i v({\bf r}_i) =
E_k - \frac{(\hat{T}+\hat{U})\Psi_k}{\Psi_k},
\label{invert}
\ee
which shows that any eigenstate $\Psi_k$ (and thus in particular the
ground state $\Psi_0=\Psi[n_0]$) determines the potential operator $\hat{V}$ up 
to an additive constant, the corresponding eigenenergy. As a consequence, 
the explicit reference to the potential $v$ in the energy functional $E_v[n]$ 
is not necessary, and one can rewrite the ground-state energy as
\be
\fbox{$
E_0 = E[n_0] = \la \Psi[n_0]|\hat{T}+\hat{U}+\hat{V}[n_0]|\Psi[n_0] \ra.$}
\ee
Another consequence is that $n_0$ now does determine not only the GS
wave function but the complete Hamiltonian (the operators $\hat{T}$ and 
$\hat{U}$ are fixed), and thus all excited states, too:
\be
\fbox{$
\Psi_k({\bf r}_1,{\bf r}_2 \ldots, {\bf r}_N) = 
\Psi_k[n_0],$}
\ee
where $k$ labels the entire spectrum of the many-body Hamiltonian $\hat{H}$.

\subsection{Complications: $N$ and $v$-representability of densities,
and nonuniqueness of potentials}
\label{complications}

Originally the fourth statement was considered to be as sound as the 
other three. However, it has become clear very recently, as a consequence of 
work of H.~Eschrig and W.~Pickett \cite{nonunep} and, independently, of 
the author with G.~Vignale \cite{nonun1,nonun2}, that there are
significant exceptions to it. In fact, the fourth substatement holds only
when one formulates DFT exclusively in terms of the charge density, as
we have done up to this point. It does not hold when one works with
spin densities (spin-DFT) or current densities (current-DFT).\footnote{In
Section \ref{extensions} we will briefly discuss these formulations of DFT.}
In these (and some other) cases the densities still determine the wave 
function, but they do not uniquely determine the corresponding potentials. 
This so-called {\it nonuniqueness problem} has been discovered only recently, 
and its consequences are now beginning to be explored 
\cite{stronghk,nonunep,nonun1,nonun2,argaman,nikitas,kohnnonun,carstennonun}. 
It is clear, however, that the fourth
substatement is, from a practical point of view, the least important of the
four, and most applications of DFT do not have to be reconsidered as a
consequence of its eventual failure. 
(But some do: see Refs.~\cite{nonun1,nonun2} for examples.)

Another conceptual problem with the HK theorem, much better known and more
studied than nonuniqueness, is representability. To understand what
representability is about, consider the following two questions:
(i) How does one know, given an arbitrary function $n\r$, that this function 
can be represented in the form (\ref{densdef}), i.e., that it is a density 
arising from an antisymmetric $N$-body wave function 
$\Psi({\bf r}_1\ldots{\bf r}_N)$?
(ii) How does one know, given a function that can be written in the form 
(\ref{densdef}), that this density is a ground-state density of a local
potential $v\r$? The first of these questions is known as the 
$N$-representability
problem, the second is called $v$-representability. Note that these are quite 
important questions: if one should find, for example, in a numerical 
calculation, a 
minimum of $E_{v}[n]$ that is not $N$-representable, then this minimum is not 
the physically acceptable solution to the problem at hand. Luckily, the 
$N$-representability problem of the single-particle density has been solved: 
any nonnegative function can be written in terms of some antisymmetric
$\Psi({\bf r}_1,{\bf r}_2 \ldots, {\bf r}_N)$ in the form (\ref{densdef}) 
\cite{gilbert,harriman}.

No similarly general solution is known for the $v$-representability problem.
(The HK theorem only guarantees that there cannot be {\em more than one} 
potential for each density, but does not exclude the possibility that there 
is {\em less than one}, i.e., zero, potentials capable of producing that 
density.) It is known that in {\em discretized} systems every density is 
{\em ensemble} $v$-representable, which means that a local potential with a 
degenerate ground state can always be found, such that the density $n\r$ can 
be written as linear combination of the densities arising from each of the 
degenerate ground states \cite{chayes,ullrichkohn,lammert}.
It is not known if one of the two restrictions (`discretized systems', and 
`ensemble') can be relaxed in general (yielding `in continuum systems' and 
`pure-state' respectively), but it is known that one may not relax both: 
there are densities in continuum systems that are not representable by a 
single nondegenerate antisymmetric function that is ground state of a local 
potential $v\r$ \cite{dftbook,chayes,ullrichkohn,lammert}.
In any case, the constrained search algorithm of Levy and Lieb shows that
$v$-representability in an interacting system is not required for the 
proof of the HK theorem. For the related question of simultaneous 
$v$-representability in a noninteracting system, which appears in the 
context of the Kohn-Sham formulation of DFT, see footnotes 34 and 35.

\subsection{A preview of practical DFT}
\label{preview}

After these abstract considerations let us now consider one way in which
one can make practical use of DFT. Assume we have specified our system
(i.e., $v\r$ is known). Assume further that we have reliable approximations for
$U[n]$ and $T[n]$. In principle, all one has to do then is to minimize the 
sum of kinetic, interaction and potential energies
\be
E_v[n] = T[n] + U[n] + V[n] = T[n] + U[n] + \int d^3r \, n\r v\r
\label{minimize}
\ee
with respect to $n\r$. 
The minimizing function $n_0\r$ is the system's
GS charge density and the value $E_{v,0}=E_v[n_0]$ is the GS energy.
Assume now that $v\r$ depends on a parameter $a$. This can be, for example,
the lattice constant in a solid or the angle between two atoms in a molecule.
Calculation of $E_{v,0}$ for many values of $a$ allows one to plot the curve 
$E_{v,0}(a)$ and to find the value of $a$ that minimizes it. This value, 
$a_0$, is 
the GS lattice constant or angle. In this way one can calculate quantities
like molecular geometries and sizes, lattice constants, unit cell volumes,
charge distributions, total energies, etc. By looking at the change of
$E_{v,0}(a)$ with $a$ one can, moreover, calculate compressibilities, phonon
spectra and bulk moduli (in solids) and vibrational frequencies (in molecules).
By comparing the total energy of a composite system (e.g., a molecule)
with that of its constituent systems (e.g., individual atoms) one obtains
dissociation energies. By calculating the total energy for systems with one
electron more or less one obtains electron affinities and ionization 
energies.\footnote{Electron affinities are typically harder to obtain than 
ionization energies, because within the local-density and 
generalized-gradient approximations the $N+1$'st electron is too weakly bound 
or even unbound: the asymptotic effective potential obtained from these
approximations decays exponentially, and not as $1/r$, i.e., it approaches
zero so fast that binding of negative ions is strongly suppressed. 
Self-interaction corrections or other fully nonlocal functionals are needed to
improve this behaviour.}
By appealing to the Hellman-Feynman theorem one can calculate 
forces on atoms from the derivative of the total energy with respect to the
nuclear coordinates. All this follows from DFT without having to solve
the many-body Schr\"odinger equation and without having to make a
single-particle approximation. For brief comments on the most useful additional 
possibility to also calculate single-particle band structures see 
Secs.~\ref{kseqs} and \ref{hartreeetal}.

In theory it should be possible to calculate {\it all} observables, since the
HK theorem guarantees that they are all functionals of $n_0\r$. In practice,
one does not know how to do this explicitly. Another problem is that the
minimization of $E_{v}[n]$ is, in general, a tough numerical problem on its
own. And, moreover, one needs reliable approximations for $T[n]$ and $U[n]$
to begin with. In the next section, on the Kohn-Sham equations, we will see 
one widely used method for solving these problems. Before looking at that, 
however, it is worthwhile to recall an older, but still occasionally
useful, alternative: the Thomas-Fermi approximation.

In this approximation one sets
\be
U[n] \approx U_H[n] = 
\frac{q^2}{2} \int d^3r \int d^3r' \frac{n\r n\rp}{|{\bf r}-{\bf r'}|},
\label{uhartree}
\ee
i.e., approximates the full interaction energy by the Hartree energy, the
electrostatic interaction energy of the charge distribution $n\r$. One
further approximates, initially,
\be
T[n] \approx T^{LDA}[n] = \int d^3r \, t^{hom}(n\r),
\label{tlda}
\ee
where $t^{hom}(n)$ is the kinetic-energy density of a homogeneous 
interacting system with (constant) density $n$. Since it refers to interacting
electrons $t^{hom}(n)$ is not known explicitly, and further approximations
are called for. As it stands, however,
this formula is already a first example of a local-density
approximation (LDA). In this type of approximation one imagines the
real inhomogeneous system (with density $n\r$ in potential $v\r$) to be
decomposed in small cells in each of which $n\r$ and $v\r$ are approximately
constant. In each cell (i.e., locally) one can then use the per-volume
energy of a homogeneous system to approximate the contribution of the 
cell to the real inhomogeneous one. Making the cells infinitesimally small 
and summing over all of them yields Eq.~(\ref{tlda}). 

For a noninteracting system (specified by subscript $s$, for `single-particle')
the function $t^{hom}_s(n)$ is known explicitly,
$t^{hom}_s(n) = 3\hbar^2(3\pi^2)^{2/3} n^{5/3}/(10m)$ 
(see also Sec.~\ref{local}). This is exploited to further approximate
\be
T[n] \approx T^{LDA}[n] \approx T_s^{LDA}[n] = \int d^3r \, t_s^{hom}(n\r),
\label{tfapproxt}
\ee
where $T_s^{LDA}[n]$ is the local-density approximation to $T_s[n]$, the 
kinetic energy of noninteracting electrons of density $n$. Equivalently, it 
may be considered the noninteracting version of $T^{LDA}[n]$. (The quantity 
$T_s[n]$ will reappear below, in discussing the Kohn-Sham equations.)
The Thomas-Fermi approximation\footnote{The Thomas-Fermi approximation
for screening, discussed in many books on solid-state physics, is obtained
by minimizing $E^{TF}[n]$ with respect to $n$ and linearizing the 
resulting relation between $v\r$ and $n\r$. It thus involves one more
approximation (the linearization) compared to what is called the 
Thomas-Fermi approximation in DFT \cite{anatf}. In two dimensions no
linearization is required and both become equivalent \cite{anatf}.} 
then consists in combining (\ref{uhartree})
and (\ref{tfapproxt}) and writing
\be
E[n]=T[n] + U[n] + V[n] \approx E^{TF}[n] = T_s^{LDA}[n] + U_H[n] + V[n].
\label{tfapprox}
\ee
A major defect of the Thomas-Fermi approximation is that within it 
molecules are unstable: the energy of a set of isolated atoms is lower
than that of the bound molecule. This fundamental deficiency, and the
lack of accuracy resulting from neglect of correlations in (\ref{uhartree})
and from using the local approximation (\ref{tfapproxt}) for the kinetic 
energy, limit the practical use of
the Thomas-Fermi approximation in its own right. However, it is found a
most useful starting point for a large body of work on improved approximations
in chemistry and physics \cite{march1,march2}. More recent approximations
for $T[n]$ can be found, e.g., in Refs.~\cite{teter,madden,carter}, in
the context of orbital-free DFT. The extension of the local-density concept 
to the exchange-correlation energy is at the heart of many modern density 
functionals (see Sec.~\ref{local}).

\subsection{From wave functions to density functionals via Green's functions
and density matrices}
\label{wavefunctions}

It is a fundamental postulate of quantum mechanics that the wave function 
contains all possible information about a system in a pure state at zero 
temperature, whereas at nonzero temperature this information is contained
in the density matrix of quantum statistical mechanics. 
Normally, this is much more information that one can handle: for a system
with $N=100$ particles the many-body wave function is an extremely
complicated function of $300$ spatial and $100$ spin\footnote{To keep the
notation simple, spin labels are either ignored or condensed into a
common variable $x:=({\bf r}s)$ in most of this text. They will only be put 
back explicitly in discussing
spin-density-functional theory, in Sec.~\ref{extensions}.} variables that
would be impossible to manipulate algebraically or to extract any information
from, even if it were possible to calculate it in the first place.
For this reason one searches for less complicated objects to formulate the
theory. Such objects should contain the experimentally relevant information,
such as energies, densities, etc., but do not need to contain explicit
information about the coordinates of every single particle.
One class of such objects are {\it Green's functions}, which are described in
the next subsection, and another are {\em reduced density matrices}, described 
in the subsection \ref{densmat}. Their relation to the wave function and the 
density is summarized in Fig.~\ref{fig1}.

\subsubsection{Green's functions}
\label{greensf}

Readers with no prior knowledge of (or no interest in) Green's functions 
should skip this subsection, which is not necessary for understanding the 
following sections.

In mathematics one
usually defines the Green's function of a linear operator ${\cal L}$ via
$[z-{\cal L}(r)]G(x,x';z)=\delta(x-x')$, where $\delta(x-x')$ is Dirac's
delta function. For a single quantum particle in potential $v\r$ one has,
for example,
\be
\left[E+\frac{\hbar^2\nabla^2}{2m} - v\r \right] G^{(0)}({\bf r},{\bf r}';E)
=\hbar\delta({\bf r}-{\bf r}').
\label{freeg}
\ee
Many applications of such single-particle Green's functions are discussed
in Ref.~\cite{economou}. In many-body physics it is useful to also
introduce more complicated Green's functions. In an interacting system
the single-particle Green's function is modified by the presence of the
interaction between the particles.\footnote{Note that expressions like
`two-particle operator' and `single-particle Green's function' refer to the
number of particles involved in the definition of the operator (two in the
case of an interaction, one for a potential energy, etc.), not to the total
number of particles present in the system.}
In general it now satisfies the equation\footnote{When energy is conserved, 
i.e., the Hamiltonian does not depend
on time, $G({\bf r},t;{\bf r}',t')$ depends on time only via the difference
$t-t'$ and can be written as $G({\bf r},{\bf r}';t-t')$.
By Fourier transformation with respect to $t-t'$ one then
passes from $G({\bf r},{\bf r}';t-t')$ to $G({\bf r},{\bf r}';E)$ of
Eq.~(\ref{freeg}).}
\bea
\left[i\hbar \frac{\partial }{\partial t}
+ \frac{\hbar^2\nabla^2}{2m} - v\r \right]
G({\bf r},t;{\bf r}',t') = \hbar \delta({\bf r}-{\bf r}') \delta(t-t')
\nonumber \\
- i\int d^3x\, U({\bf r}-{\bf x})
G^{(2)}({\bf r}t,{\bf x}t;{\bf r}'t',{\bf x}t^+),
\label{greeneqofm}
\eea
where $G^{(2)}({\bf r}t,{\bf x}t;{\bf r}'t',{\bf x}t^+)$ is the
two-particle Green's function \cite{economou,mbt}. Only for noninteracting
systems ($U=0$) is $G({\bf r},t;{\bf r}',t')$ a Green's function in the
mathematical sense of the word. In terms of $G({\bf r},t;{\bf r}',t')$ one can 
explicitly express the expectation value of any single-body operator (such 
as the potential, the kinetic energy, the particle density, etc.), and also 
that of certain two-particle operators, such as the Hamiltonian in the presence
of particle-particle interactions.\footnote{For arbitrary two-particle 
operators one needs the full two-particle Green's function $G^{(2)}$.}

One way to obtain the single-particle Green's function is via solution of
what is called Dyson's equation \cite{economou,mbt,szabo},
\bea
G({\bf r},t;{\bf r}',t') = G^{(0)}({\bf r},t;{\bf r}',t')
\nonumber \\
+ \int d^3x \int d^3x' \int d^3\tau \int d^3\tau' \,
G^{(0)}({\bf r},t;{\bf x},\tau)
\Sigma({\bf x},\tau,{\bf x}',\tau')
G({\bf x}',\tau';{\bf r}',t'),
\label{dyson}
\eea
where $\Sigma$ is known as the irreducible self energy \cite{economou,mbt,szabo}
and $G^{(0)}$ is the Green's function in the absence of any interaction.
This equation (which we will not attempt to solve here) has a characteristic
property that we will meet again when we study the (much simpler)
Kohn-Sham and Hartree-Fock equations, in Sec.~\ref{singlebody}:
the integral on the right-hand side,
which determines $G$ on the left-hand side, depends on $G$ itself.
The mathematical problem posed by this equation is thus {\it nonlinear}.
We will return to such nonlinearity when we discuss self-consistent solution
of the Kohn-Sham equation. The quantity $\Sigma$ will appear again in
Sec.~\ref{hartreeetal} when we discuss the
meaning of the eigenvalues of the Kohn-Sham equation.

The single-particle Green's function is related to the irreducible self energy
by Dyson's equation (\ref{dyson}) and to the two-particle Green's function by 
the equation of motion (\ref{greeneqofm}).
It can also be related to the $xc$ potential of DFT by the Sham-Schl\"uter 
equation \cite{ss}
\bea
\int d^3r' v_{xc}\r \int \omega 
G_s({\bf r},{\bf r}';\omega)G({\bf r}',{\bf r};\omega)
=
\nonumber \\
\int d^3r' \int d^3r'' \int \omega 
G_s({\bf r},{\bf r}';\omega) \Sigma_{xc}({\bf r}',{\bf r}'';\omega) 
G(y{\bf r}'',{\bf r};\omega),
\label{shscheq}
\eea
where $G_s$ is the Green's function of noninteracting particles with
density $n\r$ (i.e., the Green's function of the Kohn-Sham 
equation, see Sec. \ref{kseqs}), and 
$\Sigma_{xc}({\bf r}',{\bf r}'';\omega) = 
\Sigma({\bf r}',{\bf r}'';\omega) -\delta({\bf r}'-{\bf r}'')v_H\rp$ 
represents all contributions to the full irreducible self energy beyond
the Hartree potential.

A proper discussion of $\Sigma$ and $G$ requires a formalism known as
second quantization \cite{economou,mbt} and usually proceeds via introduction
of Feynman diagrams. These developments are beyond the scope of the present
overview. A related concept, density matrices, on the other hand, can be 
discussed easily. The next section is devoted to a brief description of some 
important density matrices.

\subsubsection{Density matrices}
\label{densmat}

For a general quantum system at temperature $T$, the density operator
in a canonical ensemble is defined as
\be
\hat{\gamma} = \frac{\exp^{-\beta \hat{H}}}{Tr[\exp^{-\beta \hat{H}}]},
\label{densop}
\ee
where $Tr[\cdot]$ is the trace and $\beta =1/(k_BT)$.
Standard textbooks on statistical physics show how this operator is obtained
in other ensembles, and how it is used to calculate thermal and quantum
expectation values. Here we focus on the relation to density-functional
theory. To this end we write $\hat{\gamma}$ in the energy representation as
\be
\hat{\gamma} = 
\frac{\sum_i \exp^{-\beta E_i}|\Psi_i\ra\la \Psi_i|}{\sum_i\exp^{-\beta E_i}},
\ee
where $| \Psi_i\ra$ is eigenfunction of $\hat{H}$, and the sum is over 
the entire spectrum of the system, each state being weighted by its Boltzmann 
weight $\exp^{-\beta E_i}$. At zero temperature only 
the ground-state contributes to the sums, so that 
\be
\hat{\gamma} = | \Psi_i\ra\la \Psi_i|.
\ee
The coordinate-space matrix element of this operator for an $N$-particle 
system is 
\bea
\la x_1,x_2,..x_N | \hat{\gamma}| x_1',x_2',..x_N' \ra
= \Psi(x_1,x_2,..x_N)^* \Psi(x_1',x_2',..x_N')
\nonumber \\
=:\gamma(x_1,x_2,..x_N;x_1',x_2',..x_N'),
\eea
which shows the connection between the density matrix and the wave function.
(We use the usual abbreviation $x={\bf r}s$ for space and spin coordinates.)
The expectation value of a general $N$-particle operator $\hat{O}$ is obtained 
from $O=\la \hat{O} \ra = \int dx_1 \int dx_2...\int dx_N 
\Psi(x_1,x_2,..x_N)^* \hat{O} \Psi(x_1,x_2,..x_N)$, which for multiplicative 
operators becomes
\be
\la \hat{O} \ra = \int dx_1 \int dx_2...\int dx_N\, 
\hat{O}\, \gamma(x_1,x_2,..x_N;x_1,x_2,..x_N)
\label{opexp}
\ee
and involves only the function $\gamma(x_1,x_2,..x_N;x_1,x_2,..x_N)$, which 
is the diagonal element of the matrix $\gamma$. Most operators
we encounter in quantum mechanics are one or two-particle operators and can 
be calculated from reduced density matrices, that depend on less than 
$2N$ variables.\footnote{Just as for Green's functions, expressions like 
`two-particle operator' and `two-particle density matrix' refer to the
number of particles involved in the definition of the operator (two in the
case of an interaction, one for a potential energy, etc.), not to the total
number of particles present in the system.} The reduced two-particle density 
matrix is defined as
\bea
\gamma_2(x_1,x_2;x_1',x_2')= 
\nonumber \\
\frac{N(N-1)}{2} \int dx_3 \int dx_4...\int dx_N
\gamma(x_1,x_2,x_3,x_4,..x_N;x_1',x_2',x_3,x_4,..x_N),
\label{gamma2def}
\eea
where $N(N-1)/2$ is a convenient normalization factor. This density matrix
determines the expectation value of the particle-particle interaction, of
static correlation and response functions, of the $xc$ hole, and some related 
quantities. The pair-correlation function $g(x,x')$, e.g., is obtained from 
the diagonal element of $\gamma_2(x_1,x_2;x_1',x_2')$ according to
$\gamma_2(x_1,x_2;x_1,x_2)=:n(x_1)n(x_2)g(x,x')$.

Similarly, the single-particle density matrix is defined as
\bea
\gamma(x_1,x_1') = 
\nonumber \\
N \int dx_2 \int dx_3 \int dx_4 ...\int dx_N
\gamma(x_1,x_2,x_3,x_4,..x_N;x_1',x_2,x_3,x_4,..x_N)
\nonumber \\
=N \int dx_2 \int dx_3 \int dx_4 ...\int dx_N 
\Psi^*(x_1,x_2,x_3,..,x_N)
\Psi(x_1',x_2,x_3,..,x_N).
\label{gammadef2}
\eea
The structure of reduced density matrices is quite simple: all coordinates 
that $\gamma$ does not depend upon are set equal in $\Psi$ and $\Psi^*$, and
integrated over.  The single-particle density matrix can also be considered 
the time-independent form of the single-particle Green's function, since it
can alternatively be obtained from
\be
\gamma(x,x') = -i\lim_{t'\to t}G(x,x';t-t').
\label{gammadef1}
\ee

In the special case that the wave function $\Psi$ is a Slater determinant,
i.e., the wave function of $N$ noninteracting fermions, the 
single-particle density matrix can be written in terms of the orbitals 
comprising the determinant, as
\be
\gamma(x,x')= \sum_j \phi_j^*(x)\phi_j(x'),
\ee
which is known as the Dirac (or Dirac-Fock) density matrix.

The usefulness of the single-particle density matrix becomes apparent when
we consider how one would calculate the expectation value of a
multiplicative single-particle operator $\hat{A} =\sum_i^N a({\bf r}_i) $
(such as the potential $\hat{V}= \sum_i^N v({\bf r}_i)$):
\bea
\la \hat{A} \ra = \int dx_1 \ldots \int dx_N \,
\Psi^*(x_1,x_2,..,x_N)
\left[ \sum_i^N a(x_i) \right]
\Psi(x_1,x_2,..,x_N)
\\
= N \int dx_1 \ldots \int dx_N \,
\Psi^*(x_1,x_2,..,x_N)
a(x_1)
\Psi(x_1,x_2,..,x_N)
\\
= \int dx\, a(x) \gamma(x,x),
\label{opev}
\eea
which is a special case of Eq.~(\ref{opexp}). 
The second line follows from the first by exploiting that the fermionic
wave function $\Psi$ changes sign upon interchange of two of its arguments.
The last equation implies that if one knows $\gamma(x,x)$ one
can calculate the expectation value of any multiplicative single-particle
operator in terms of it, regardless of the number of particles present in
the system.\footnote{For nonmultiplicative single-particle operators (such 
as the kinetic energy, which contains a derivative) one requires the full
single-particle matrix $\gamma(x,x')$ and not only $\gamma(x,x)$.}  
The simplification is enormous, and reduced density matrices 
are very popular in, e.g., computational chemistry for precisely this reason. 
More details are given in, e.g., Ref.~\cite{parryang}. The full density 
operator, Eq.~(\ref{densop}), on the other hand, is the central quantity of
quantum statistical mechanics.

It is not possible to express expectation values of two-particle operators, 
such as the interaction itself, or the full Hamiltonian (i.e., the total 
energy), explicitly in terms of the single-particle density matrix
$\gamma\rrp$. For this purpose one requires the two-particle density matrix. 
This situation is to be contrasted with that of the single-particle Green's 
function, for which one knows how to express the expectation values of 
$\hat{U}$ and $\hat{H}$ \cite{economou,mbt}. 
Apparently, some information has gotten lost
in passing from $G$ to $\gamma$. This can also be seen very clearly from
Eq.~(\ref{gammadef1}), which shows that information on the dynamics of the
system, which is contained in $G$, is erased in the definition of 
$\gamma\rrp$. Explicit information on the static properties of the system is
contained in the $N$-particle density matrix, but as seen from
(\ref{gamma2def}) and (\ref{gammadef2}), a large part of this 
information is also lost ('integrated out') in passing from
the $N$-particle density matrix to the reduced two- or one-particle density
matrices.

Apparently even less information is contained in the particle 
density\footnote{A quantitative estimate of how much less information is
{\em apparently} contained in the density than in the wave function is
given in footnote 6.}
$n\r$, which is obtained by summing the diagonal element of $\gamma(x,x')$
over the spin variable,
\be
n\r = \sum_s \gamma({\bf r}s,{\bf r}s). 
\ee
This equation follows immediately from comparing (\ref{densdef}) with 
(\ref{gammadef2}). We can define an alternative density operator, $\hat{n}$,
by requiring that the same equation must also be obtained by substituting 
$\hat{n}\r$ into Eq.~(\ref{opev}), which holds for any single-particle
operator. This requirement implies that 
$\hat{n}\r=\sum_i^N \delta({\bf r}-{\bf r}_i)$.\footnote{The expectation value 
of $\hat{n}$ is the particle density, and therefore $\hat{n}$ is often also
called the density operator. This concept must not be confused with any of 
the various density matrices or the density operator of statistical physics, 
Eq.~(\ref{densop}).}

\begin{figure}[t]
\centering
\includegraphics[height=65mm,width=65mm,angle=0]{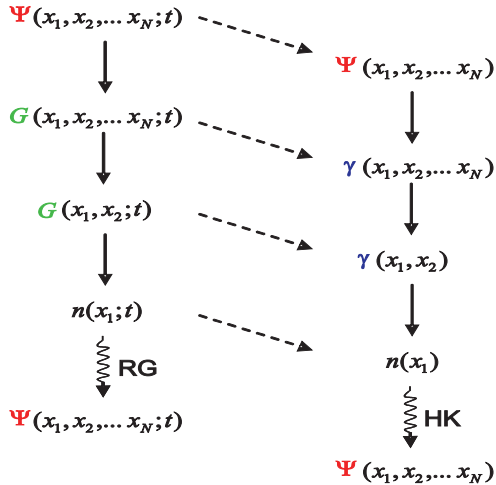}
\caption {\label{fig1}
Information on the time-and-space dependent wave function
$\Psi(x_1,x_2 \ldots, x_N,t)$ is built into
Green's functions, and on the time-independent wave function into density
matrices. Integrating out degrees of freedom reduces the $N$-particle
Green's function and $N$-particle density matrix to the one-particle
quantities $G(x_1,x_2;t)$ and $\gamma(x_1,x_2)$ described in the main
text. The diagonal element of the one-particle density matrix is the
ordinary charge density --- the central quantity in DFT. The Hohenberg-Kohn
theorem and its time-dependent generalization (the Runge-Gross theorem)
guarantee that the densities contain exactly the same information as the
wave functions.
}
\end{figure}

The particle density is an even simpler function than $\gamma(x,x')$: it 
depends on one set of coordinates $x$ only, it can easily be visualized
as a three-dimensional charge distribution, and it is directly accessible
in experiments.
These advantages, however, seem to be more than compensated by the fact that
one has integrated out an enormous amount of specific
information about the system in going from wave functions to Green's functions,
and on to density matrices and finally the density itself. This process
is illustrated in Fig. \ref{fig1}.

The great surprise of density-functional theory is that in fact no information
has been lost at all, at least as long as one considers the system only
in its ground state: according to the Hohenberg-Kohn theorem the ground-state 
density $n_0(x)$ completely determines the ground-state wave function
$\Psi_0(x_1,x_2 \ldots, x_N)$.\footnote{The Runge-Gross theorem, which forms 
the basis of time-dependent DFT \cite{rungegross}, similarly guarantees that
the time-dependent density contains the same information as the time-dependent
wave function.}

Hence, in the ground state, a function of one variable is equivalent to
a function of $N$ variables! This property
shows that we have only integrated out {\it explicit} information on our way
from wave functions via Green's functions and density matrices to densities.
{\it Implicitly} all the information that was contained in the ground-state
wave function is still contained in the ground-state density. Part of the
art of practical DFT is how to get this implicit information out, once one
has obtained the density!

\section{DFT as an effective single-body theory: The Kohn-Sham equations}
\label{singlebody}
\hspace{10mm}

Density-functional theory can be implemented in many ways. The minimization of
an explicit energy functional, discussed up to this point, is not normally
the most efficient among them. Much more widely used is the Kohn-Sham 
approach. Interestingly, this approach owes its success and popularity
partly to the fact that it does not exclusively work in terms of
the particle (or charge) density, but brings a special kind of wave functions
(single-particle orbitals) back into the game. As a consequence DFT then looks
formally like a single-particle theory, although many-body effects are
still included via the so-called exchange-correlation functional. We will 
now see in some detail how this is done.

\subsection{Exchange-correlation energy: definition, interpretation and
exact properties}
\label{xcenergy}

\subsubsection{Exchange-correlation energy}

The Thomas-Fermi approximation (\ref{tfapproxt}) for $T[n]$ is not very good.
A more accurate scheme for treating the kinetic-energy functional of 
interacting electrons, $T[n]$, is based on decomposing it into one part that 
represents the kinetic energy of noninteracting particles of density $n$, 
i.e., the quantity called above $T_s[n]$, and one that represents the
remainder, denoted $T_c[n]$ (the subscripts $s$ and $c$ stand for 
`single-particle' and `correlation', respectively).\footnote{
$T_s$ is defined as the expectation value of the kinetic-energy operator
$\hat{T}$ with the Slater determinant arising from density $n$, i.e.,
$T_s[n]=\la \Phi[n]| \hat{T} | \Phi[n] \ra$. Similarly, the full kinetic
energy is defined as $T[n]=\la \Psi[n]| \hat{T} | \Psi[n] \ra$. 
All consequences of antisymmetrization (i.e., exchange) are described by
employing a determinantal wave function in defining $T_s$. Hence, $T_c$, the
difference between $T_s$ and $T$ is a pure correlation effect.}
\be
T[n] = T_s[n] + T_c[n].
\ee
$T_s[n]$ is not known exactly as a functional of $n$ [and using the LDA to
approximate it leads one back to the Thomas-Fermi approximation 
(\ref{tfapproxt})], but it is
easily expressed in terms of the single-particle orbitals $\phi_i\r$
of a noninteracting system with density $n$, as
\be
T_s[n] = -\frac{\hbar^2}{2m} \sum_i^N \int d^3r\, 
\phi_i^*\r \nabla^2 \phi_i\r,
\label{torbital}
\ee
because for noninteracting particles the total kinetic energy is just the sum
of the individual kinetic energies. Since all $\phi_i\r$ are functionals of 
$n$, this expression for $T_s$ is an explicit orbital functional but an  
implicit density functional, $T_s[n] = T_s[\{\phi_i[n]\}]$, where the
notation indicates that $T_s$ depends on the full set of occupied orbitals
$\phi_i$, each of which is a functional of $n$.
Other such orbital functionals will be discussed in Sec.~\ref{approx}.

We now rewrite the exact energy functional as
\be
\fbox{$
E[n] = T[n] + U[n] + V[n] =
T_s[\{\phi_i[n]\}] + U_H[n] + E_{xc}[n] + V[n],$}
\label{excdef}
\ee
where by definition $E_{xc}$ contains the differences $T-T_s$ (i.e. $T_c$)
and $U-U_H$. This definition shows that a significant part of the correlation
energy $E_c$ is due to the difference $T_c$ between the noninteracting and
the interacting kinetic energies. Unlike Eq.~(\ref{tfapprox}), 
Eq.~(\ref{excdef}) is formally exact, but of course $E_{xc}$ is
unknown --- although the HK theorem guarantees that it is a density functional.
This functional, $E_{xc}[n]$, is called the {\it exchange-correlation} 
(xc) energy. It is often decomposed as $E_{xc}=E_x+E_c$, where $E_x$ is due
to the Pauli principle (exchange energy) and $E_c$ is due to correlations.
($T_c$ is then a part of $E_c$.) The exchange energy can be written 
explicitly in terms of the single-particle orbitals as\footnote{This differs 
from the exchange energy used in Hartree-Fock theory only in the substitution 
of Hartree-Fock orbitals $\phi_i^{HF}\r$ by Kohn-Sham orbitals $\phi_i\r$.} 
\be
\fbox{$
E_x[\{\phi_i[n]\}] =
-\frac{q^2}{2} \sum_{jk} \int d^3r \int d^3r' \,
\frac{\phi^*_j\r \phi^*_k\rp \phi_j\rp \phi_k\r}{|{\bf r}-{\bf r}'|},$}
\label{fock}
\ee
which is known as the Fock term, but no general exact expression in terms 
of the density is known. 

\subsubsection{Different perspectives on the correlation energy}
\label{corrint}

For the correlation energy no general explicit
expression is known, neither in terms of orbitals nor densities.
Different ways to understand correlations are described below.

{\em Correlation energy: variational approach.}
A simple way to understand the origin of correlation is to recall that
the Hartree energy is obtained in a variational calculation in which the 
many-body wave function is approximated as a product of single-particle 
orbitals. Use of an antisymmetrized product (a Slater determinant) produces
the Hartree and the exchange energy \cite{mbt,szabo}. The correlation energy 
is then defined as the difference between the full ground-state energy 
(obtained with the correct many-body wave function) and the one obtained from 
the (Hartree-Fock or Kohn-Sham\footnote{The Hartree-Fock
and the Kohn-Sham Slater determinants are not identical, since they are
composed of different single-particle orbitals, and thus the definition of
exchange and correlation energy in DFT and in conventional quantum chemistry
is slightly different \cite{grabogross}.}) Slater determinant. 
Since it arises from a more general trial wave function than a single Slater 
determinant, correlation cannot raise the total energy, and $E_c[n]\leq 0$.
Since a Slater determinant is itself more general than a simple product
we also have $E_{x}\leq 0$, and thus the upper bound\footnote{A lower bound 
is provided by the Lieb-Oxford formula, given in 
Eq.~(\ref{lobound}).} $E_{xc}[n] \leq 0$.

{\em Correlation energy: probabilistic approach.}
Recalling the quantum mechanical interpretation of the wave function as
a probability amplitude, we see that a product form of the many-body wave
function corresponds to treating the probability amplitude of the 
many-electron system as a product of the probability amplitudes of 
individual electrons (the orbitals). Mathematically, the probability of 
a composed event is only equal to the probability of the individual events 
if the individual events are independent (i.e., uncorrelated). Physically,
this means that the electrons described by the product wave function are 
independent.\footnote{Correlation is a mathematical concept describing the
fact that certain events are not independent. It can be defined also in 
classical physics, and in applications of statistics to other fields than 
physics. Exchange is due to the indistinguishability of particles, and a true
quantum phenomenon, without any analogue in classical physics.}
Such wave functions thus neglect the fact that, as a consequence 
of the Coulomb interaction, the electrons try to avoid each other. The
correlation energy is simply the additional energy lowering obtained 
in a real system due to the mutual avoidance of the interacting electrons.
One way to characterize a strongly correlated system is to define correlations
as strong when $E_c$ is comparable in magnitude to, or larger than, other 
energy contributions, such as $E_H$ or $T_s$. (In weakly correlated systems 
$E_c$ normally is several orders of magnitude smaller.)\footnote{Other 
characterizations of strongly correlated systems are to compare the width of 
the conduction band in
a solid with the kinetic energy (if the band width is smaller, correlations
are strong), or the quasiparticle energies $\tilde{\eps}_i$ with the Kohn-Sham
eigenvalues $\eps_i$ (if both are similar, correlations are weak, see
footnote 37), or the derivative discontinuity $\Delta_{xc}$, defined in
Eq.~(\ref{xcdisc}), with the Kohn-Sham energy gap (if the former is comparable
to or larger than the latter, correlations are strong). (The meaning of
$\tilde{\eps}_i$, $\eps_i$ and $\Delta_{xc}$ is explained below.) No
universally applicable definition of `strong correlations' seems to exist.} 

{\em Correlation energy: beyond mean-field approach.}
A rather different (but equivalent) way to understand correlation is to 
consider the following alternative form of the operator representing the 
Coulomb interaction, equivalent to Eq.~(\ref{clbdef}),
\be
\hat{U} = \frac{q^2}{2}\int d^3r \int d^3r' \, 
\frac{\hat{n}\r \hat{n}\rp -\hat{n}\r\delta({\bf r}-{\bf r'})}
{|{\bf r}-{\bf r'}|},
\label{clbdef2}
\ee
in which the operator character is carried only by the density operators
$\hat{n}$ (in occupation number representation), and the term with the 
delta function subtracts out the interaction of a charge with itself
(cf., e.g., the appendix of Ref.~\cite{pinesnozieres} for a simple
derivation of this form of $\hat{U}$).
The expectation value of this operator, $U=\la \Psi|\hat{U}|\Psi \ra$, 
involves the expectation value of a product of density operators,
$\la \Psi|\hat{n}\r\hat{n}\rp|\Psi \ra$. In the Hartree term (\ref{uhartree}), 
on the other hand, this expectation value of a product is replaced by a 
product of expectation values, each of the form 
$n\r=\la \Psi|\hat{n}\r|\Psi \ra$.
This replacement amounts to a mean-field approximation, which neglects quantum 
fluctuations\footnote{At finite temperature there are also thermal 
fluctuations. To properly include these one must use a finite-temperature 
formulation of DFT \cite{mermin}. See also the contribution of B.~L.~Gyorffy 
et al. in Ref.~\cite{asi95} for DFT treatment of various types of fluctuations.}
about the expectation values: by writing $\hat{n} = n + \delta\hat{n}_{fluc}$ 
and substituting in Eq.~(\ref{clbdef2}) we see that the difference between 
$\la \Psi|\hat{U}|\Psi \ra$ and the Hartree term (\ref{uhartree}) is entirely 
due to the fluctuations $\delta\hat{n}_{fluc}$ and the self-interaction 
correction to the Hartree term. Quantum fluctuations about the expectation
value are thus at the origin of quantum correlations between interacting
particles.

{\em Correlation energy: holes.}
The fact that both exchange and correlation tend to keep electrons apart,
has given rise to the concept of an $xc$ hole, $n_{xc}\rrp$, describing 
the reduction of probability for encountering an electron at ${\bf r'}$,
given one at ${\bf r}$. The $xc$ energy can be written as a Hartree-like
interaction between the charge distribution $n\r$ and the $xc$ hole
$n_{xc}\rrp=n_x\rrp+n_c\rrp$,
\be 
E_{xc}[n]={q^2\over 2} \int d^3r \int d^3r'\, 
\frac{n\r n_{xc}\rrp}{|{\bf r}-{\bf r'}|},
\ee
which defines $n_{xc}$.
The exchange component $E_x[n]$ of the exact exchange-correlation 
functional describes the energy lowering due to antisymmetrization (i.e., 
the tendency of like-spin electrons to avoid each other). It gives rise to the 
exchange hole $n_x\rrp$, which obeys the sum rule $\int d^3 r'\, n_x\rrp =-1$. 
The correlation component $E_c[n]$ accounts for the additional energy lowering 
arising because electrons with opposite spins also avoid each other. 
The resulting correlation hole integrates to zero, so that the total
$xc$ hole satisfies $\int d^3 r'\, n_{xc}\rrp =-1$. The $xc$ hole can also 
be written as $n_{xc}\rrp=n({\bf r'})(\bar{g}[n]\rrp-1)$, where $\bar{g}$ 
is the average of the pair-correlation function $g\rrp$, mentioned in 
Sec.~\ref{densmat}, over all values of the particle-particle interaction, 
from zero (KS system) to $\la\hat{U}\ra$ (interacting system). This average 
is simply expressed in terms of the coupling constant $\alpha$ as
$\bar{g}\rrp=\int_0^1 g_\alpha\rrp d\alpha$. For the Coulomb interaction,
$\alpha = e^2$, i.e., the square of the electron charge \cite{dftbook,parryang}.

\subsubsection{Exact properties}

Clearly $E_c$ is an enormously complex object, and DFT would be of little use 
if one had to know it exactly for making calculations. 
The practical advantage of writing $E[n]$ in the form 
Eq.~(\ref{excdef}) is that the unknown functional $E_{xc}[n]$ is typically
much smaller than the known terms $T_s$, $U_H$ and $V$. One can thus hope
that reasonably simple approximations for $E_{xc}[n]$ provide useful
results for $E[n]$. Some successful approximations are discussed in
Sec.~\ref{approx}. Exact properties, such as the sum rule
$\int d^3 r' n_{xc}\rrp =-1$, described in the preceding section, are most 
valuable guides in the construction of approximations to $E_{xc}[n]$. 

Among the known properties of this functional are the coordinate scaling 
conditions first obtained by Levy and Perdew \cite{scaling}
\bea
E_x[n_\lambda] &=& \lambda E_x[n] \label{cooscaling1}\\
E_c[n_\lambda] &>& \lambda E_c[n] \hspace{1cm} {\rm for}\, \lambda >1
\label{cooscaling2} \\
E_c[n_\lambda] &<& \lambda E_c[n] \hspace{1cm} {\rm for}\, \lambda <1,
\label{cooscaling3} 
\eea
where $n_\lambda\r=\lambda^3 n(\lambda {\bf r})$ is a scaled density 
integrating to total particle number $N$. 

Another important property of the exact functional is the one-electron limit
\bea
E_c[n^{(1)}] &\equiv& 0 \\
E_x[n^{(1)}] &\equiv& - E_H[n^{(1)}],
\eea
where $n^{(1)}$ is a one-electron density. These latter two conditions, which 
are satisfied within the Hartree-Fock approximation, but not by standard 
local-density and gradient-dependent functionals, ensure that there is no 
artificial self-interaction of one electron with itself. 

The Lieb-Oxford bound \cite{lieboxford,chanhandy},
\be
E_x[n] \geq E_{xc}[n] \geq -1.68 e^2 \int d^3r \, n\r^{4/3},
\label{lobound}
\ee
establishes a lower bound on the $xc$ energy, and is satisfied by LDA and
many (but not all) GGAs.

One of the most intriguing properties of the exact functional, which has
resisted all attempts of describing it in local or semilocal approximations,
is the derivative discontinuity of the $xc$ functional with respect to the 
total particle number
\cite{ss,lp1,lp2},
\be
\left.\frac{\delta E_{xc}[n]}{\delta n\r}\right|_{N+\delta}
- \left.\frac{\delta E_{xc}[n]}{\delta n\r}\right|_{N-\delta}
=
v_{xc}^+\r - v_{xc}^-\r
= \Delta_{xc},
\label{xcdisc}
\ee
where $\delta$ is an infinitesimal shift of the electron number $N$,
and $\Delta_{xc}$ is a system-dependent, but ${\bf r}$-independent shift
of the $xc$ potential $v_{xc}\r$ as it passes from the electron-poor to the
electron-rich side of integer $N$. The noninteracting kinetic-energy
functional has a similar discontinuity, given by
\be
\left.\frac{\delta T_s[n]}{\delta n\r}\right|_{N+\delta}
- \left.\frac{\delta T_s[n]}{\delta n\r}\right|_{N-\delta}
=\eps_{N+1}-\eps_{N} = \Delta_{KS},
\label{tsdisc}
\ee
where $\eps_N$ and $\eps_{N+1}$ are the Kohn-Sham (KS) single-particle 
energies of the highest occupied and lowest unoccupied eigenstate.
The meaning of these KS eigenvalues is discussed in the paragraphs
following Eq.~(\ref{Vdef}) and illustrated in Fig. \ref{fig2}. 
In the chemistry literature these 
are called the HOMO (highest occupied molecular orbital) and LUMO
(lowest unoccupied molecular orbital), respectively. The kinetic-energy 
discontinuity is thus simply the KS single-particle gap $\Delta_{KS}$, or
HOMO-LUMO gap, whereas the $xc$ discontinuity $\Delta_{xc}$ is a many-body 
effect. The true fundamental gap $\Delta = E(N+1)+E(N-1)-2 E(N)$ is 
the discontinuity of the total ground-state energy functional 
\cite{dftbook,ss,lp1,lp2},
\be
\fbox{$
\Delta = \left.\frac{\delta E[n]}{\delta n\r}\right|_{N+\delta}
- \left.\frac{\delta E[n]}{\delta n\r}\right|_{N-\delta}
= \Delta_{KS} + \Delta_{xc}.$}
\label{gap}
\ee
Since all terms in $E$ other than $E_{xc}$ and $T_s$ are continuous
functionals of $n\r$, the fundamental gap is the sum of the KS gap and the
$xc$ discontinuity. Standard density functionals (LDA and GGA) 
predict $\Delta_{xc}=0$, and thus often underestimate the fundamental gap.
The fundamental and KS gaps are also illustrated in Fig. \ref{fig2}.

All these properties serve as constraints or guides in the construction
of approximations for the functionals $E_x[n]$ and $E_c[n]$. Many other
similar properties are known. A useful overview of scaling properties is
the contribution of M.~Levy in Ref.~\cite{asi95}. 

\subsection{Kohn-Sham equations}
\label{kseqs}

\subsubsection{Derivation of the Kohn-Sham equations}

Since $T_s$ is now written as an orbital functional one cannot directly
minimize Eq.~(\ref{excdef}) with respect to $n$. Instead, one 
commonly employs a scheme suggested by Kohn and Sham \cite{ks} for performing
the minimization indirectly. This scheme starts by writing the minimization
as
\be
0=\frac{\delta E[n]}{\delta n\r} = \frac{\delta T_s[n]}{\delta n\r}
+ \frac{\delta V[n]}{\delta n\r}
+ \frac{\delta U_H[n]}{\delta n\r} + \frac{\delta E_{xc}[n]}{\delta n\r}
= \frac{\delta T_s[n]}{\delta n\r} + v\r + v_H\r + v_{xc}\r.
\label{minim}
\ee
As a consequence of Eq.~(\ref{vdef}), $\delta V/\delta n = v\r$, the 
`external' potential the electrons move in.\footnote{This potential is called 
`external' because it is external to the electron system and not generated
self-consistently from the electron-electron interaction, as $v_H$ and 
$v_{xc}$. It comprises the lattice potential and any additional truly external 
field applied to the system as a whole.} The term $\delta U_H/\delta n$
simply yields the Hartree potential, introduced in Eq.~(\ref{hartreepotl}). 
For the term $\delta E_{xc}/\delta n$, which can only be calculated
explicitly once an approximation for $E_{xc}$ has been chosen, one commonly
writes $v_{xc}$. By means of the Sham-Schl\" uter equation (\ref{shscheq}),
$v_{xc}$ is related to the irreducible self energy $\Sigma$, introduced in 
Eq.~(\ref{dyson}) \cite{ss}.

Consider now a system of noninteracting particles moving in the
potential $v_s\r$. For this system the minimization condition is simply
\be
0=\frac{\delta E_s[n]}{\delta n\r} = 
\frac{\delta T_s[n]}{\delta n\r} + \frac{\delta V_s[n]}{\delta n\r}
= \frac{\delta T_s[n]}{\delta n\r} + v_s\r,
\ee
since there are no Hartree and $xc$ terms in the absence of interactions.
The density solving this Euler equation is $n_s\r$.
Comparing this with Eq.~(\ref{minim}) we find that both minimizations
have the same solution $n_s\r \equiv n\r$, if $v_s$ is chosen to be
\be
\fbox{$
v_s\r = v\r + v_H\r + v_{xc}\r.$}
\label{vsdef}
\ee
Consequently, one can calculate the density of the interacting (many-body) 
system in potential $v\r$, described by a many-body Schr\"odinger equation
of the form (\ref{mbse}), by solving the equations of a noninteracting 
(single-body) system in potential $v_s\r$.\footnote{The question whether such
a potential $v_s\r$ always exists in the mathematical sense is called the
noninteracting $v$-representability problem. It is known that every 
interacting ensemble $v$-representable density is also noninteracting ensemble 
$v$-representable, but, as mentioned in Sec. \ref{hktheorem}, only in 
discretized systems has it been proven that all densities are interacting 
ensemble $v$-representable. It is not known if interacting 
ensemble-representable densities may be noninteracting pure-state 
representable (i.e, by a single determinant), which would be convenient 
(but is not necessary) for Kohn-Sham calculations.}

In particular, the Schr\"odinger equation of this auxiliary system,
\be
\fbox{$
\left[ -\frac{\hbar^2 \nabla^2}{2m} + v_s\r \right] \phi_i\r = 
\eps_i \phi_i\r,$}
\label{ksequation}
\ee
yields orbitals that reproduce the density $n\r$ of the original system
(these are the same orbitals employed in Eq.~(\ref{torbital})),
\be
\fbox{$
n\r \equiv n_s\r = \sum_i^N f_i\,|\phi_i\r|^2,
$}
\label{ncalc}
\ee
where $f_i$ is the occupation of the $i$'th orbital.\footnote{Normally,
the occupation numbers $f_i$ follows an {\em Aufbau} principle (Fermi 
statistics) with $f_i=1$ for $i<N$, $f_i=0$ for $i>N$, and $0\leq f_i \leq 1$ 
for $i=N$ (i.e., at most the highest occupied orbital can have fractional 
occupation). Some densities that are not noninteracting $v$-representable
by a single ground-state Slater determinant, may still be obtained from a 
single determinant if one uses occupation numbers $f_i$ that leave holes below
the HOMO (the Fermi energy in a metal), so that $f_i\neq 1$ even for some
$i<N$ \cite{perdewlevy}, but this is not guaranteed to describe all
possible densities. Alternatively (see Sec. \ref{hktheorem} and footnote 34)
a Kohn-Sham equation may be set up in terms of ensembles of determinants.
This guarantees noninteracting $v$-representability for all densities 
that are interacting ensemble $v$-representable.
For practical KS calculations, the most important consequence of the fact
that not every arbitrary density is guaranteed to be noninteracting 
$v$-representable is that the Kohn-Sham selfconsistency cycle is not 
guaranteed to converge.} 
Eqs.~(\ref{vsdef}) to (\ref{ncalc}) are the celebrated Kohn-Sham (KS) 
equations. They replace the problem of minimizing $E[n]$ by that of solving a 
noninteracting Schr\"odinger equation. (Recall that the minimization of
$E[n]$ originally replaced the problem of solving the many-body
Schr\"odinger equation!)

Since both $v_H$ and $v_{xc}$ depend on $n$, which depends on the $\phi_i$,
which in turn depend on $v_s$, the problem of solving the KS equations
is a nonlinear one, just as is the one of solving the (much more complicated)
Dyson equation (\ref{dyson}). The usual way of solving such problems is to
start with an initial guess for $n\r$, calculate the corresponding $v_s\r$,
and then solve the differential equation (\ref{ksequation}) for the $\phi_i$.
From these one calculates a new density, using (\ref{ncalc}), and starts
again. The process is repeated until it converges. The technical name for 
this procedure is `self-consistency cycle'. Different convergence
criteria (such as convergence in the energy, the density, or some observable 
calculated from these) and various convergence-accelerating algorithms
(such as mixing of old and new effective potentials) are in common use. 
Only rarely it requires more than a few dozen iterations to achieve 
convergence, and even rarer are cases where convergence seems unattainable,
i.e., a self-consistent solution of the KS equation cannot be found. 

Once one has a converged solution $n_0$, one can calculate the total energy
from Eq.~(\ref{excdef}) or, equivalently and more conveniently, 
from\footnote{All terms on the right-hand side of (\ref{etot2})
except for the first, involving the sum of the single-particle energies, are
sometimes known as double-counting corrections, in analogy to a similar
equation valid within Hartree-Fock theory.}
\be
\fbox{$
E_0=\sum_i^N \eps_i 
-\frac{q^2}{2}\int d^3r \int d^3r'\, \frac{n_0\r n_0\rp}{|{\bf r}-{\bf r'}|}
-\int d^3r\, v_{xc}\r n_0\r + E_{xc}[n_0].$}
\label{etot2}  
\ee
Equation (\ref{etot2}) follows from writing $V[n]$ in (\ref{excdef}) by means
of (\ref{vsdef}) as
\bea
V[n]=\int d^3r \, v\r n\r = \int d^3r \, [v_s\r - v_H\r - v_{xc}\r] n\r
\\
= V_s[n] - \int d^3r \, [v_H\r + v_{xc}\r] n\r,
\label{Vdef}
\eea 
and identifying the energy of the noninteracting (Kohn-Sham) system 
as $E_s= \sum_i^N \eps_i = T_s+V_s$.

\subsubsection{The eigenvalues of the Kohn-Sham equation}
\label{eigenvalues}

Equation (\ref{etot2}) shows that $E_0$ is not simply the sum\footnote{The
difference between $E_0$ and $\sum_i^N \eps_i$ is due to particle-particle
interactions. The additional terms on the right-hand side of (\ref{etot2})
give mathematical meaning to the common statement that {\em the whole is 
more than the sum of its parts}. If $E_0$ can be written approximately as 
$\sum_i^N \tilde{\eps_i}$ (where the $\tilde{\eps_i}$ are not the same as 
the KS eigenvalues $\eps_i$) the system can be described in terms of $N$
weakly interacting quasiparticles, each with energy $\tilde{\eps_i}$. 
Fermi-liquid theory in metals and effective-mass theory in semiconductors 
are examples of this type of approach.} of all $\eps_i$.
In fact, it should be clear from our derivation of Eq.~(\ref{ksequation})
that the $\eps_i$ are introduced as completely artificial objects: they are 
the eigenvalues of an auxiliary single-body equation whose eigenfunctions
(orbitals) yield the correct density. It is only this density that has strict 
physical meaning in the KS equations. The KS eigenvalues, on the other hand, 
in general bear only a semiquantitative resemblance with the true energy 
spectrum \cite{kohnbeckeparr}, but are not to be trusted quantitatively.

The main exception to this rule is the highest occupied KS eigenvalue. 
Denoting by $\eps_N(M)$ the $N$'th eigenvalue of a system with $M$ electrons, 
one can show rigorously that $\eps_N(N)=-I$, the negative of the first 
ionization energy of the $N$-body system, and $\eps_{N+1}(N+1)=-A$, the 
negative of the electron affinity of the same $N$-body system
\cite{lp1,almbladh,lps}. These relations
hold for the exact functional only. When calculated with an approximate 
functional of the LDA or GGA type, the highest eigenvalues usually do not 
provide good approximations to the experimental $I$ and $A$. Better results 
for these observables are obtained by calculating them as total-energy 
differences, according to $ I=E_0(N-1)-E_0(N)$ and $A=E_0(N)-E_0(N+1)$, 
where $E_0(N)$ is the ground-state energy of the $N$-body system. 
Alternatively, self-interaction corrections can be used to obtain improved      ionization energies and electron affinities from Kohn-Sham eigenvalues
\cite{sickli1}.

\begin{figure}[t]
\centering
\includegraphics[height=55mm,width=70mm,angle=0]{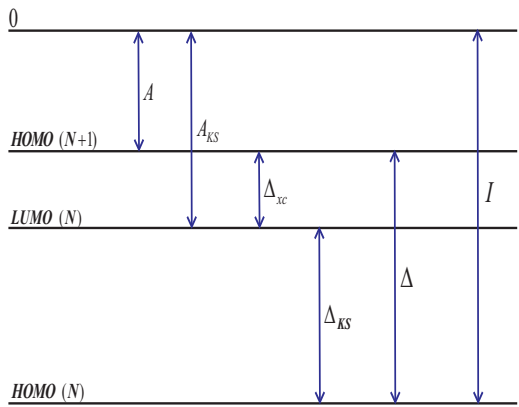}
\caption {\label{fig2}
Schematic description of some important Kohn-Sham eigenvalues
relative to the vacuum level, denoted by $0$, and their 
relation to observables. See main text for explanations.}
\end{figure}

Figure \ref{fig2} illustrates the role played by the highest occupied and 
lowest unoccupied KS eigenvalues, and their relation to observables.
For molecules, HOMO(N) is the highest-occupied molecular orbital
of the $N$-electron system, HOMO(N+1) that of the $N+1$-electron system,
and LUMO(N) the lowest unoccupied orbital of the $N$-electron system. In
solids with a gap, the HOMO and LUMO become the top of the valence band and
the bottom of the conduction band, respectively, whereas in metals they are
both identical to the Fermi level. The vertical lines indicate the Kohn-Sham
(single-particle) gap $\Delta_{KS}$, the fundamental (many-body) gap
$\Delta$, the derivative discontinuity of the $xc$ functional, $\Delta_{xc}$,
the ionization energy of the interacting $N$-electron system 
$I(N)=-\eps_N(N)$ (which is also the ionization energy of the Kohn-Sham 
system $I_{KS}(N)$), the electron affinity of the interacting $N$-electron 
system $A(N)=-\eps_{N+1}(N+1)$ and the Kohn-Sham electron affinity 
$A_{KS}(N)=-\eps_{N+1}(N)$. 

Given the auxiliary nature of the other Kohn-Sham eigenvalues, it comes as 
a great (and welcome) surprise that in many situations 
(typically characterized by the presence of fermionic quasiparticles and 
absence of strong correlations) the Kohn-Sham eigenvalues $\eps_i$ do,
empirically, provide a reasonable first approximation to the actual energy 
levels of extended systems. This approximation is behind most band-structure
calculations in solid-state physics, and often gives results that agree well 
with experimental photoemission and inverse photoemission data \cite{photoem},
but much research remains to be done before it is clear to what extent such 
conclusions can be generalized, and how situations in which the KS eigenvalues 
are good starting points for approximating the true excitation spectrum are 
to be characterized microscopically \cite{harrison,umrigar}.\footnote{Several 
more rigorous approaches to excited states in DFT, which do not require the KS 
eigenvalues to have physical meaning, are mentioned in Sec.~\ref{extensions}.}

Most band-structure calculations in solid-state physics are actually
calculations of the KS eigenvalues $\eps_i$.\footnote{A computationally 
more expensive, but more reliable, alternative is provided by the
GW approximation \cite{gw1,gw2,rubio}.}
This simplification has proved enormously
successful, but when one uses it one must be aware of the fact that one is 
taking the auxiliary single-body equation (\ref{ksequation}) literally
as an approximation to the many-body Schr\"odinger equation. DFT, practiced
in this mode, is not a rigorous many-body theory anymore, but a
mean-field theory (albeit one with a very sophisticated mean field $v_s\r$). 

The energy gap obtained in such band-structure calculations is the one
called HOMO-LUMO gap in molecular calculations, i.e., the difference 
between the energies of the highest occupied and the lowest unoccupied 
single-particle states. Neglect of the derivative discontinuity $\Delta_{xc}$,
defined in Eq.~(\ref{xcdisc}), by standard local and semilocal $xc$ functionals
leads to an underestimate of the gap (the so-called `band-gap problem'),
which is most severe in transition-metal oxides and other strongly
correlated systems. Self-interaction corrections provide a partial
remedy for this problem \cite{svanesic,temmermansic,naturesic,sickli2}.

\subsubsection{Hartree, Hartree-Fock and Dyson equations}
\label{hartreeetal}

A partial justification for the interpretation of the KS eigenvalues as 
starting point for approximations to quasi-particle energies, common in 
band-structure calculations, can be given by comparing the KS equation 
with other self-consistent equations of many-body physics.
Among the simplest such equations are the Hartree equation
\be
\left[-\frac{\hbar^2\nabla^2}{2m}+v\r+v_H\r\right]\phi^H_i\r
=\eps^H_i\phi^H_i\r,
\ee
and the Hartree-Fock (HF) equation
\be
\left[-\frac{\hbar^2\nabla^2}{2m}+v\r+v_H\r\right]\phi^{HF}_i\r
-q^2\int d^3r'\,\frac{\gamma({\bf r},{\bf r'})}{|{\bf r}-{\bf r'}|}\phi^{HF}_i\rp
=\eps^{HF}_i\phi^{HF}_i\r,
\label{hartreefock}
\ee
where $\gamma\rrp$ is the density matrix of Eq.~(\ref{gammadef2}).
It is a fact known as Koopman's theorem \cite{szabo}
that the HF eigenvalues $\eps^{HF}_i$ can be interpreted as unrelaxed 
electron-removal energies (i.e., ionization energies of the $i$'th electron,
neglecting reorganization of the remaining electrons after removal).
As mentioned above, in DFT only the highest occupied eigenvalue corresponds
to an ionization energy, but unlike in HF this energy includes relaxation
effects.

The KS equation (\ref{ksequation}) includes both exchange and correlation via 
the multiplicative operator $v_{xc}$. Both exchange and correlation are
normally approximated in DFT,\footnote{A possibility to treat exchange exactly 
in DFT is offered by the OEP method discussed in Sec.~\ref{nonlocal}.}
whereas HF accounts for exchange exactly, through the integral operator 
containing $\gamma\rrp$, but neglects correlation completely. 
In practise DFT results are typically at least as good as HF ones and often
comparable to much more sophisticated correlated methods --- and the KS 
equations are much easier to solve than the HF equations.\footnote{This is 
due to the integral operator in the HF equations.} 

All three single-particle equations, Hartree, Hartree-Fock and Kohn-Sham
can also be interpreted as approximations to Dyson's equation (\ref{dyson}),
which can be rewritten as \cite{mbt} 
\be
\left(-\frac{\hbar^2\nabla^2}{2m} + v\r \right) \psi_k\r
+ \int d^3r'\, \Sigma({\bf r},{\bf r}',E_k)\psi_k({\bf r}') = E_k \psi_k\r,
\label{dyson2}
\ee
where $\Sigma$ is the irreducible self energy introduced in Eq.~(\ref{dyson}).
The $E_k$ appearing in this equation are the true (quasi-)electron addition 
and removal energies of the many-body system. Needless to say, it is much 
more complicated to solve this equation than the HF or KS equations. It is 
also much harder to find useful approximations for $\Sigma$ than for 
$v_{xc}$.\footnote{The GW approximation \cite{gw1,gw2,rubio}, mentioned in 
footnote 39, is one such approximation for $\Sigma$, but in actual 
implementations of it one usually takes DFT-KS results as an input.}
Obviously, the KS equation employs a local, energy-independent potential
$v_s$ in place of the nonlocal, energy-dependent operator $\Sigma$.
Whenever this is a good approximation, the $\eps_i$ are also a good
approximation to the $E_k$.

The interpretation of the KS equation (\ref{ksequation}) as an approximation 
to Eq.~(\ref{dyson2})
is suggestive and useful, but certainly not necessary for DFT to work:
if the KS equations are only used to obtain the density, and all other 
observables, such as total energies, are calculated from this density, 
then the KS equations in themselves are not an approximation at all, but 
simply a very useful mathematical tool.

\subsection{Basis functions}
\label{basisfunctions}

In practice, numerical solution of the KS differential equation
(\ref{ksequation}) typically proceeds by expanding the KS orbitals in a 
suitable set of basis functions and solving the resulting secular equation 
for the coefficients in this expansion and/or for the eigenvalues for which it 
has a solution. The construction of suitable basis functions is a major 
enterprise within electronic-structure theory (with relevance far beyond DFT), 
and the following lines do little more than explaining some acronyms often used
in this field. 

In physics much is known about the construction of basis functions for solids
due to decades of experience with band-structure calculations. This includes 
many calculations that predate the widespread use of DFT in physics. There is a
fundamental dichotomy between methods that work with fixed basis functions that
do not depend on energy, and methods that employ energy-dependent basis 
functions.
Fixed basis functions are used e.g., in plane-wave expansions, tight-binding
or LCAO (linear combination of atomic orbitals) approximations, or the 
OPW (orthogonalized plane wave) method. Examples for methods using 
energy-dependent functions are the APW (augmented plane wave)
or KKR (Korringa-Kohn-Rostoker) approaches. This distinction became less
clear-cut with the introduction of `linear methods' \cite{andersenprl},
in which energy-dependent basis functions are linearized (Taylor expanded)
around some fixed reference energy. The most widely used methods for 
solving the Kohn-Sham equation in solid-state physics, LMTO (linear
muffin tin orbitals) and LAPW (linear augmented plane waves), are of this
latter type \cite{andersen}. Development of better basis functions is
an ongoing enterprise \cite{nlmto1,nlmto2}.

The situation is quite similar in chemistry. Due to decades of experience with 
Hartree-Fock and CI calculations much is known about the construction of basis 
functions that are suitable for molecules. Almost all of this continues to hold
in DFT --- a fact that has greatly contributed to the recent popularity of
DFT in chemistry.
Chemical basis functions are classified with respect to their behaviour
as a function of the radial coordinate into Slater type orbitals (STOs),
which decay exponentially far from the origin, and 
Gaussian type orbitals (GTOs), which have a gaussian behaviour.
STOs more closely resemble the true behaviour of atomic wave functions
[in particular the cusp condition of Eq.~(\ref{kato})],
but GTOs are easier to handle numerically because the product of two GTOs
located at different atoms is another GTO located in between, whereas the
product of two STOs is not an STO. 
The so-called `contracted basis functions', in which STO basis functions are 
reexpanded in a small number of GTOs, represent a compromise between the 
accuracy of STOs and the convenience of GTOs. The most common methods for 
solving the Kohn-Sham equations in quantum chemistry are of this type
\cite{rmp2,szabo}. Very accurate basis functions for chemical purposes 
have been constructed by Dunning \cite{dunning} and, more recently, by
da Silva and collaborators \cite{alberico1,alberico2}.
More details on the development of suitable basis functions 
can be found, e.g., in these references and Ref.~\cite{szabo}.

A very popular approach to larger systems in DFT, in particular solids, is 
based on the concept of a pseudopotential (PP). The idea behind the PP is that
chemical binding in molecules and solids is dominated by the outer (valence) 
electrons of each atom. The inner (core) electrons retain, to a good 
approximation, an atomic-like configuration, and their orbitals do not change 
much if the atom is put in a different environment. Hence, it is possible to
approximately account for the core electrons in a solid or a large molecule 
by means of an atomic calculation, leaving only the valence density to be
determined self-consistently for the system of interest.

In the original 
Kohn-Sham equation the effective potential $v_s[n]=v_{ext}+v_H[n]+v_{xc}[n]$ 
is determined by the full electronic density $n\r$, and the self-consistent
solutions are single-particle orbitals reproducing this density. In the
PP approach the Hartree and $xc$ terms in $v_s[n]$ are evaluated only 
for the valence density $n_v$, and the core electrons are accounted for by
replacing the external potential $v_{ext}$ by a pseudopotential $v_{ext}^{PP}$.
Hence $v_s^{PP}[n_v]=v_{ext}^{PP}+v_H[n_v]+v_{xc}[n_v]$.\footnote{Note that 
the {\em effective} potential $v_s$ is a way to deal with the electron-electron
interaction. The {\em pseudopotential} is a way to deal with the density of 
the core electrons. Both potentials can be profitably used together, but
are conceptually different.}
The PP $v_{ext}^{PP}$ is determined in two steps. First, one determines, 
in an auxiliary atomic calculation, an effective PP, $v_{s}^{PP}$, 
such that for a suitably chosen atomic reference configuration the 
single-particle orbitals resulting from $v_{s}^{PP}$ agree --- outside a 
cut-off radius $r_c$ separating the core from the valence region --- with 
the valence orbitals obtained from the all-electron KS equation for the same 
atom. As a consequence, the valence densities $n_v^{at}$ obtained from the 
atomic KS and the atomic PP equation are the same. Next, one subtracts the 
atomic valence contributions $v_H[n_v^{at}]$ and $v_{xc}[n_v^{at}]$ from 
$v_{s}^{PP}[n_v^{at}]$ to obtain the external PP $v_{ext}^{PP}$,\footnote{This
external PP is also called the unscreened PP, and the subtraction of 
$v_H[n_v^{at}]$ and $v_{xc}[n_v^{at}]$ from $v_{s}^{PP}[n_v^{at}]$ is called 
the 'unscreening of the atomic PP'. It can only be done exactly 
for the Hartree term, because the contributions of valence and core densities 
are not additive in the $xc$ potential (which is a nonlinear functional of 
the total density).} which is then used in the molecular or 
solid-state calculation, together with $v_H[n_v]$ and $v_{xc}[n_v]$ taken at 
the proper valence densities for these systems.

The way $v_s^{PP}$ is generated from the atomic calculation is not unique.
Common pseudopotentials are generated following the prescription of, e.g.,
Bachelet, Hamann and Schl\"uter \cite{ppbhs}, Kleinman and Bylander
\cite{ppkb}, Vanderbilt \cite{ppv} or Troullier and Martins \cite{pptm}.
Useful reviews are Refs.~\cite{ppp,ppppl,ppptaaj}.
The pseudopotential approach is very convenient because it reduces
the number of electrons treated explicitly, making it possible to
perform density-functional calculations on systems with tens of
thousands of electrons. Moreover, the pseudopotentials $v_{ext}^{PP}$ are 
much smoother than the bare nuclear potentials $v_{ext}$. The remaining 
valence electrons are thus well described by plane-wave basis sets.

\begin{figure}
\centering
\includegraphics[scale=0.5]{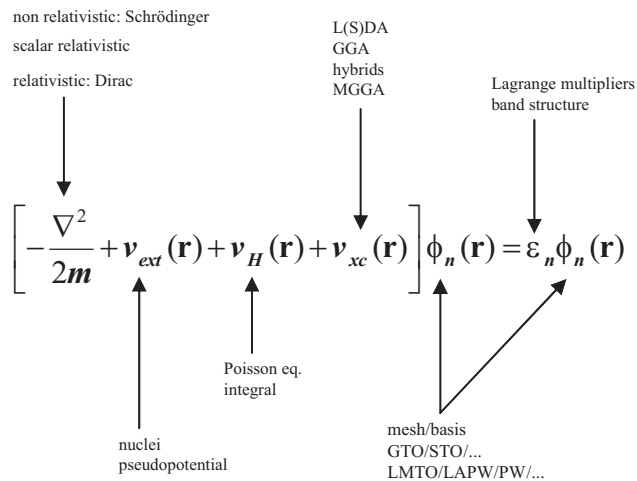}
\caption {\label{fig3}
Some of the choices made in a Kohn-Sham calculation. The treatment can be
nonrelativistic (based on Schr\" odinger's equation), scalar relativistic 
(using the relativistic kinetic-energy operator and other simple relativistic 
corrections, but no spin-orbit coupling) or relativistic (using Dirac's 
equation, which includes also spin-orbit coupling). The core electrons can 
be treated explicitly (all electron 
calculation) or incorporated, together with $v_{ext}$, in a pseudopotential.
The Hartree potential can be obtained from integrating the charge density 
or from solving Poisson's differential equation. Many choices are available
for the $xc$ potential. The eigenvalues can be considered mere Lagrange 
multipliers or interpreted as zero-order approximations to the actual energy 
spectrum. The eigenfunctions can similarly be considered auxiliary functions 
generating the density, or interpreted as zero-order approximations to 
quasi-particle wave functions. Solution of the KS equation can proceed on a 
numerical mesh, or by expansion of the eigenfunctions in basis functions. 
Many types of suitable basis functions exist. For every new problem a
suitable combination of choices must be made, and all possibilities continue
to be useful and to be actively explored in physics and chemistry.}
\end{figure}

Some of the choices one has to make in a practical Kohn-Sham calculation are
illustrated schematically in Fig.~\ref{fig3}.

\section{Making DFT practical: Approximations}
\label{approx}
\hspace{10mm}

There are basically three distinct types of approximations involved in a
DFT calculation. One is conceptual, and concerns the interpretation of
KS eigenvalues and orbitals as physical energies and wave functions. 
This approximation is optional --- if one does not want to make it one simply 
does not attach meaning to the eigenvalues of Eq.~(\ref{ksequation}). The 
pros and cons of this procedure were discussed in Secs.~\ref{eigenvalues} and
\ref{hartreeetal}.
The second type of approximation is numerical, and concerns methods for
actually solving the differential equation (\ref{ksequation}).
A main aspect here is the selection of suitable basis functions, briefly
discussed in Sec.~\ref{basisfunctions}.
The third type of approximation involves constructing an expression for
the unknown $xc$ functional $E_{xc}[n]$, which contains all many-body
aspects of the problem [cf. Eq.~(\ref{excdef})]. It is with this type of
approximation that we are concerned in the present section.

This chapter is intended to give the reader an idea of what types of 
functionals exist, and to describe what their main features are, separately for
local functionals (TF, LDA and $X\alpha$; Sec.~\ref{local}), semilocal, or 
gradient-dependent, functionals (GEA and GGA; Sec.~\ref{semilocal}), and 
nonlocal functionals (hybrids, orbital functionals such as meta-GGAs, EXX 
and SIC, and integral-dependent functionals such as ADA; Sec.~\ref{nonlocal}). 
This chapter does deal only most superficially with the actual 
construction of these functionals. For more details on functional construction 
and testing the reader is referred to the reviews [5-19] or to the original 
papers cited below. Sticking to the bird's-eye philosophy of this overview
I have also refrained from including numerical data on the performance
of each functional --- extensive comparisons of a wide variety of functionals
can be found in Refs.~[5-19] and in the original literature cited below.

\subsection{Local functionals: LDA}
\label{local}

Historically (and in many applications also practically) the most important
type of approximation is the local-density approximation (LDA). To understand
the concept of an LDA recall first how the noninteracting kinetic energy 
$T_s[n]$ is treated
in the Thomas-Fermi approximation: In a homogeneous system one knows that, 
per volume\footnote{The change from a capital $T$ to a lower-case $t$ is 
commonly used to indicate quantities per volume.}
\be
t_s^{hom}(n)= \frac{3\hbar^2}{10m}(3 \pi^2)^{2/3} n^{5/3}
\ee
where $n=const$. In an inhomogeneous system, with $n=n\r$, one approximates
locally 
\be
t_s\r\approx t_s^{hom}(n\r)=\frac{3\hbar^2}{10m}(3\pi^2)^{2/3}n\r^{5/3}
\ee
and obtains the full kinetic energy by integration over all space
\be
T_s^{LDA}[n] = \int d^3r \, t_s^{hom}(n\r) = 
\frac{3\hbar^2}{10m}(3\pi^2)^{2/3} \int d^3r \, n\r^{5/3}.
\ee
For the kinetic energy the approximation $T_s[n]\approx T_s^{LDA}[n]$ is
much inferior to the exact treatment of $T_s$ in terms of orbitals, offered
by the Kohn-Sham equations, but the LDA concept turned out to be highly
useful for another component of the total energy (\ref{excdef}), the 
exchange-correlation energy $E_{xc}[n]$. For the exchange energy $E_x[n]$
the procedure is very simple, since the per-volume exchange energy of the
homogeneous electron liquid is known exactly \cite{dftbook,parryang},
\be
e_x^{hom}(n)=-\frac{3q^2}{4}\left(\frac{3}{\pi}\right)^{1/3} n^{4/3},
\ee 
so that
\be
\fbox{$
E_x^{LDA}[n] = -\frac{3q^2}{4}\left(\frac{3}{\pi}\right)^{1/3}
\int d^3r \, n\r^{4/3}.$}
\ee
This is the LDA for $E_x$.\footnote{If one adds this term to the Thomas-Fermi 
expression (\ref{tfapprox}) one obtains the so-called Thomas-Fermi-Dirac 
approximation to $E[n]$. It one multiplies it with an adjustable parameter
$\alpha$ one obtains the so-called $X\alpha$ approximation to $E_{xc}[n]$.
These approximations are not much used today in DFT.} 

For the correlation energy $E_c[n]$ the situation is more complicated since
$e_c^{hom}(n)$ is not known exactly: the determination of the correlation
energy of a homogeneous interacting electron system (an electron liquid) 
is already a difficult
many-body problem on its own! Early approximate expressions for $e_c^{hom}(n)$
were based on applying perturbation theory (e.g. the random-phase 
approximation) to this problem \cite{vbh,gl}. These approximations became
outdated with the advent of highly precise Quantum Monte Carlo (QMC) 
calculations
for the electron liquid, by Ceperley and Alder \cite{ceperley}. Modern
expressions for $e_c^{hom}(n)$ \cite{vwn,pz,pw} are parametrizations 
of these data. These expressions are implemented in most standard
DFT program packages and in typical applications give almost identical
results. On the other hand, the earlier parametrizations of the LDA, based 
on perturbation theory \cite{vbh,gl}, can occasionally deviate substantially 
from the QMC ones, and are better avoided.

Independently of the parametrization, the LDA for $E_{xc}[n]$ formally
consists in\footnote{Sometimes one uses the per-particle instead of the 
per-volume energy of the homogeneous system in writing the LDA. Since
the conversion factor between both is the number of particles per volume,
i.e., the density, an additional $n\r$ then appears under the integrals
in (\ref{ldadef}) and also contributes to (\ref{ldapotl}).}
\be
\fbox{$
E_{xc}[n] \approx E_{xc}^{LDA}[n] = \int d^3r \, 
e_{xc}^{hom}(n)|_{n\to n\r} 
= \int d^3r \, e_{xc}^{hom}(n\r),$}
\label{ldadef}
\ee
where $e_{xc}^{hom} = e_{x}^{hom}+e_c^{hom}$.
The corresponding $xc$ potential is simply
\be
\left.
v_{xc}^{LDA}[n]\r=
\frac{\partial e_{xc}^{hom}(n)}{\partial n}\right|_{n \to n\r}.
\label{ldapotl}
\ee
This approximation for $E_{xc}[n]$ has proved amazingly successful, even 
when applied to systems that are quite different from the electron liquid
that forms the reference system for the LDA. A partial explanation for this
success of the LDA is systematic error cancellation: typically, LDA 
underestimates $E_c$ but overestimates $E_x$, resulting in unexpectedly
good values of $E_{xc}$. This error cancellation is not accidental, but 
systematic, and caused by the fact that for any density the LDA $xc$ hole
satisfies the correct sum rule $\int d^3 r' n^{LDA}_{xc}\rrp =-1$
(see Sec. \ref{corrint}), which is only possible if integrated errors in
$n^{LDA}_x$ cancel with those of $n^{LDA}_c$.

For many decades the LDA has 
been applied in, e.g., calculations of band structures and total energies
in solid-state physics. In quantum chemistry it is much less popular, because
it fails to provide results that are accurate enough to permit
a quantitative discussion of the chemical bond in molecules
(so-called `chemical accuracy' requires calculations with an error of
not more than about $1$ kcal/mol $= 0.04336$ eV/particle).

At this stage it may be worthwhile to recapitulate what practical DFT
does, and where the LDA enters its conceptual structure: What real systems, 
such as atoms, molecules, clusters and solids, have in common, is that they 
are simultaneously inhomogeneous (the electrons are exposed to spatially 
varying electric fields produced by the nuclei) and interacting (the
electrons interact via the Coulomb interaction). The way density-functional
theory, in the local-density approximation, deals with this inhomogeneous
many-body problem is by decomposing it into two simpler (but still highly
nontrivial) problems: the solution of a spatially homogeneous interacting
problem (the homogeneous electron liquid) yields the uniform $xc$ energy
$e^{hom}_{xc}(n)$, and the solution of a spatially inhomogeneous 
noninteracting problem (the inhomogeneous electron gas described by the KS
equations) yields the particle density. Both steps are connected by the 
local-density potential (\ref{ldapotl}), which shows how the $xc$ 
energy of the uniform interacting system enters the equations for the
inhomogeneous noninteracting system.

The particular way
in which the inhomogeneous many-body problem is decomposed, and the various
possible improvements on the LDA, are behind the success of DFT in practical
applications of quantum mechanics to real materials. Some such improvements
on the LDA are discussed in the next two sections.

\subsection{Semilocal functionals: GEA, GGA and beyond}
\label{semilocal}

In the LDA one exploits knowledge of the density at point ${\bf r}$. 
Any real system is spatially inhomogeneous, i.e., it has a spatially varying
density $n\r$, and it would clearly be useful to also include information
on the rate of this variation in the functional. A first attempt at doing this
were the so-called `gradient-expansion approximations' (GEA). 
In this class of approximation one tries to systematically calculate
gradient-corrections of the form $|\nabla n\r|$, $|\nabla n\r|^2$, 
$\nabla^2 n\r$, etc., to the LDA. 
A famous example is the lowest-order gradient correction to the Thomas-Fermi
approximation for $T_s[n]$,
\be
T_s[n]\approx T_s^W[n] =
 T_s^{LDA}[n] + \frac{\hbar^2}{8m} \int d^3r \, \frac{|\nabla n\r|^2}{n\r}.
\ee
This second term on the right-hand side is called the Weizs\"acker 
term.\footnote{If one adds this term to the Thomas-Fermi
expression (\ref{tfapprox}) one obtains the so-called Thomas-Fermi-Weizs\"acker
approximation to $E[n]$. In a systematic gradient expansion the $8$ in the
denominator is replaced by a $72$ \cite{dftbook,parryang}.}
Similarly, in
\be
E_x[n]\approx
E_x^{GEA(2)}[n] = 
E_x^{LDA}[n] - \frac{10 q^2}{432 \pi (3 \pi^2)^{1/3}} 
\int d^3r \, \frac{|\nabla n\r|^2}{n\r^{4/3}}
\ee
the second term on the right-hand side is the lowest-order gradient 
correction\footnote{Remarkably, the form of this term is fully determined 
already by dimensional analysis: 
In $E_x^{GEA(2)}=q^2\int d^3r\, f(n,|\nabla n|^2)$ the function $f$ must
have dimensions (length)$^{-4}$. Since the dimensions of $n$ and
$|\nabla n|^2$ are (length)$^{-3}$ and (length)$^{-8}$, respectively, 
and to second order no higher powers or higher derivatives of $n$ are allowed,
the only possible combination is $f \propto |\nabla n\r|^2/n^{4/3}$.}
to $E_x^{LDA}[n]$.
In practice, the inclusion of low-order gradient corrections almost never
improves on the LDA, and often even worsens it. Higher-order corrections
(e.g., $\propto |\nabla n\r|^\alpha$ or $\propto \nabla^\beta n\r$ 
with $\alpha,\beta > 2$), on the other hand, are exceedingly difficult to
calculate, and little is known about them. 

In this situation it was a major breakthrough when it was realized, in the 
early eighties, that instead of power-series-like systematic gradient
expansions one could experiment with more general functions of $n\r$ and
$\nabla n\r$, which need not proceed order by order. Such functionals, of
the general form
\be
E_{xc}^{GGA}[n] = \int d^3r\, f(n\r,\nabla n\r),
\ee
have become known as generalized-gradient approximations (GGAs)
\cite{pw86}.

Different GGAs differ in the choice of the function $f(n,\nabla n)$.
Note that this makes different GGAs much more different from each other
than the different parametrizations of the LDA: essentially there is only
one correct expression for $e_{xc}^{hom}(n)$, and the various
parametrizations of the LDA \cite{vbh,gl,vwn,pz,pw} are merely different 
ways of writing it.
On the other hand, depending on the method of construction employed for
obtaining $f(n,\nabla n)$ one can obtain very different GGAs. 
In particular, GGAs used in quantum chemistry typically proceed by fitting 
parameters to test sets of selected molecules. On the other hand, GGAs used 
in physics tend to emphasize exact constraints.
Nowadays the most popular (and most reliable) GGAs are PBE 
(denoting the functional proposed in 1996 by Perdew, Burke and
Ernzerhof \cite{pbe}) in physics, and BLYP (denoting the combination of 
Becke's 1988 exchange functional \cite{becke} with the 1988 correlation 
functional of Lee, Yang and Parr \cite{lyp}) in chemistry. Many other 
GGA-type functionals are also available, and new ones continue to appear. 

Quite generally, current GGAs seem to give reliable results for all main
types of chemical bonds (covalent, ionic, metallic and hydrogen bridge).
For van der Waals interactions, however, common GGAs and LDA
fail.\footnote{The PBE GGA \cite{pbe} and the TPSS MGGA \cite{metaggatests} 
(see below) may be partial exceptions \cite{pbevdw,mggavdw} because they work 
reasonably well near the equilibrium distance of the van der Waals bond, 
but they recover only the short-range behaviour and do not describe correctly 
the long-range asymptotic regime of the van der Waals interaction.} 
To describe these very weak interactions several
more specialized approaches have been developed within DFT 
\cite{vdw1,vdw2,vdw3,vdw4,vdw5}.
Both in physics and in chemistry the widespread use of GGAs has lead to
major improvements as compared to LDA. `Chemical accuracy', as defined above,
has not yet been attained, but is not too far away either.
A useful collection of explicit expressions for some GGAs can be found in
the appendix of Ref.~\cite{filippi}, and more detailed discussion of some
selected GGAs and their performance is given in Ref.~\cite{kurthpropaganda}
and in the chapter of Kurth and Perdew in Refs.~\cite{joulbert,primer}.

\begin{table}[t!b!p!]
\begin{center}
\begin{tabular}{l|l}
method & -E/a.u. \\
\hline
Thomas-Fermi & 625.7 \\
Hartree-Fock & 526.818 \\
OEP (exchange only) & 526.812 \\
LDA (exchange only) & 524.517\\
LDA (VWN) & 525.946 \\
LDA (PW92) & 525.940 \\
LDA-SIC(PZ) & 528.393 \\
ADA & 527.322 \\
WDA & 528.957 \\
GGA (B88LYP) & 527.551 \\
\hline
experiment & 527.6 
% & 527.597 \\ % Clementi + rel corr. - Lamb corr.
% & 527.542 \\ % Clementi + rel corr.
% & 527.549 \\ % Caroll + rel corr.
% & 527.604 \\ % Caroll + rel corr. - Lamb corr.
% & 527.522 \\ % PZ + rel corr (13.6057)
\end{tabular}
\end{center}
\caption{\label{table1}
Ground-state energy in atomic units (1 a.u. = 1 Hartree =
2 Rydberg = $27.21 eV \hat{=} 627.5 kcal/mol$) of the $Ar$ atom ($Z=18$),
obtained with some representative density functionals and related methods.
The Hartree-Fock and OEP(exchange only) values are from Krieger et al. 
(third of Ref. \cite{krieger}), ADA and WDA values are from Gunnarsson 
et al., Ref.~\cite{adawda}, as reported in Ref.~\cite{dftbook},
and the LDA-SIC(PZ) value is from Perdew 
and Zunger, Ref.~\cite{pz}. The experimental value is based on Veillard 
and Clementi, J. Chem. Phys. {\bf 49}, 2415 (1968), and given to less 
significant digits than the calculated values, because of relativistic and 
quantum electrodynamical effects (Lamb shift) that are automatically included 
in the experimental result but not in the calculated values.}
\end{table}

No systematic attempt at comparing explicit functionals can be made here, 
but many detailed comparisons are available in the literature. For pure
illustrative purposes only, Table \ref{table1} contains ground-state
energies of the $Ar$ atom, obtained with several of the methods discussed
previously in this chapter. Footnote 7 contains additional information on
the performance of DFT for larger systems.

\subsection{Orbital functionals and other nonlocal approximations: 
hybrids, Meta-GGA, SIC, OEP, etc.}
\label{nonlocal}

In spite of these advances, the quest for more accurate functionals goes
ever on, and both in chemistry and physics various beyond-GGA functionals have
appeared. Perhaps the most popular functional in quantum chemistry\footnote{This
was written in early 2002, but at the time of revision of this text in 2006 
it is still correct.}
is B3LYP. This is a combination of the LYP GGA for correlation \cite{lyp}
with Becke's three-parameter hybrid functional B3 for exchange \cite{b3}. 
Common hybrid functionals, such as B3, mix a fraction of Hartree-Fock 
exchange into the DFT exchange functional (other mixtures are also possible). 
The construction of hybrid functional involves a certain amount of empiricism
in the choice of functionals that are mixed and in the optimization of
the weight factors given to the HF and DFT terms. Formally, this might be
considered a drawback, but in practice B3 has proven to be the most
successful exchange functional for chemical applications, in particular
when combined with the LYP GGA functional for $E_c$. More extreme
examples of this semiempirical mode of construction of functionals
are Becke's 1997 hybrid functional \cite{b10}, which contains 10 adjustable 
parameters, and the functionals of Refs.~\cite{tozer} and \cite{voorhis},
each of which contains 21 parameters. 

Another recent beyond-GGA development is the emergence of so-called
Meta-GGAs, which depend, in addition to the density and its derivatives,
also on the Kohn-Sham kinetic-energy density $\tau\r$ 
\cite{metaggatests,metagga,bmeta}
\be
\tau\r=\frac{\hbar^2}{2m}\sum_i |\nabla \phi_i\r|^2,
\label{kinden}
\ee
so that $E_{xc}$ can be written as $E_{xc}[n\r,\nabla n\r,\tau\r]$.
The additional degree of freedom provided by $\tau$ is used to satisfy
additional constraints on $E_{xc}$, such as a self-interaction-corrected
correlation functional, recovery of the fourth-order gradient expansion for 
exchange in the limit of slowly varying densities,
and a finite exchange potential at the nucleus \cite{metaggatests}.
In several recent tests \cite{metaggatests,mggavdw,adatest,mggatest1,mggatest2} 
Meta-GGAs have given favorable results, even when compared to the best 
GGAs, but the full potential of this type of approximation is only beginning 
to be explored systematically. 

As we have seen in the case of $T_s$, it can be much easier to represent a
functional in terms of single-particle orbitals than directly in terms
of the density. Such functionals are known as orbital functionals, and
Eq.~(\ref{torbital}) constitutes a simple example. Another important 
orbital-dependent functional is the exchange energy (Fock term) of 
Eq.~(\ref{fock}). The Meta-GGAs and hybrid functionals mentioned above
are also orbital functionals, because they depend on the kinetic
energy density (\ref{kinden}), and on a combination of the orbital functional
(\ref{fock}) with ordinary GGAs, respectively. 

Still another type of orbital functional is the self-interaction correction
(SIC). Most implementations of SIC make use of the expressions proposed in
Ref.~\cite{pz} (PZ-SIC),
\be
E_{xc}^{approx,SIC}[n_\ua,n_\da]= E_{xc}^{approx}[n_\ua,n_\da]-\sum_{i,\sigma}
\left(E_H[n_{i\sigma}]-E_{xc}^{approx}[n_{i\sigma},0]\right),
\ee
which subtracts, orbital by orbital, the contribution the Hartree and the $xc$
functionals would make if there was only one electron in the system.
This correction can be applied on top of any approximate density 
functional, and ensures that the resulting corrected functional satisfies
$E_{xc}^{approx,SIC}[n^{(1)},0] = -E_H[n^{(1)}]$ for a one-electron system.
The LDA is exact for a completely uniform system, and thus is self-interaction
free in this limit, but neither it nor common GGAs satisfy the requirement of
freedom from self-interaction in general, and even Meta-GGAs have a remaining
self-interaction error in their exchange part \cite{metaggatests,metagga}.
This self-interaction is particularly critical for localized states, such as
the $d$ states in transition-metal oxides. For such systems PZ-SIC has been
shown to greatly improve the uncorrected LDA \cite{svanesic,temmermansic},
but for thermochemistry PZ-SIC does not seem to be significant \cite{vydrow}.

Unfortunately the PZ-SIC approach, which minimizes the corrected energy 
functional with respect to the orbitals, does not lead to Kohn-Sham equations 
of the usual form, because the resulting effective potential is different for
each orbital. As a consequence, various specialized algorithms for minimizing
the PZ-SIC energy functional have been developed. For more details on these
algorithms and some interesting applications in solid-state physics see
Refs.~\cite{svanesic,temmermansic,naturesic}. For finite systems, PZ-SIC
has also been implemented by means of the OEP \cite{sickli1,sickli2}, which 
produces a common local potential for all orbitals, and is discussed in the 
next paragraph.
A detailed review of implementations and applications of PZ-SIC can be found
in the contribution of Temmerman et al. in Ref.~\cite{dobsonetal}.
Alternatives to the PZ-SIC formulation of Ref.~\cite{pz} have 
recently been analysed in \cite{lundin,legrand}, with a view on either
improving results obtained with PZ-SIC, or simplifying the implementation
of the correction.

Since hybrid functionals, Meta-GGAs, SIC, the Fock term and all other orbital 
functionals depend on the density only 
implicitly, via the orbitals $\phi_i[n]$, it is not possible to directly 
calculate the functional derivative $v_{xc}=\delta E_{xc}/\delta n$.
Instead one must use indirect approaches to minimize $E[n]$ and 
obtain $v_{xc}$. In the case of the kinetic-energy functional
$T_s[\{\phi_i[n]\}]$ this indirect approach is simply the Kohn-Sham scheme,
described in Sec.~\ref{singlebody}. In the case of orbital expressions
for $E_{xc}$ the corresponding indirect scheme is known as the 
optimized effective potential (OEP) \cite{krieger} or, equivalently, the
optimized-potential model (OPM) \cite{engelopm}. The minimization of the
orbital functional with respect to the density is achieved by repeated
application of the chain rule for functional derivatives,
\be
v_{xc}[n]\r= \frac{\delta E_{xc}^{orb}[\{\phi_i\}]}{\delta n\r}
=\int d^3r' \int d^3r'' \sum_i
\left[\frac{\delta E_{xc}^{orb}[\{\phi_i\}]}{\delta \phi_i\rp}
\frac{\delta \phi_i\rp}{\delta v_s\rpp}\frac{\delta v_s\rpp}{\delta n\r} 
+ c.c. \right],
\label{oepinteg}
\ee
where $E_{xc}^{orb}$ is the orbital functional (e.g., the Fock term)
and $v_s$ the KS effective potential.
Further evaluation of Eq.~(\ref{oepinteg}) gives rise to an integral equation 
that determines the $v_{xc}[n]$ belonging to the chosen orbital functional 
$E_{xc}[\{\phi_i[n]\}]$ \cite{krieger,oepreview}. As an alternative to solving 
the full OEP integral equation, Krieger, Li and Iafrate
(KLI) have proposed a simple but surprisingly accurate approximation
that greatly facilitates implementation of the OEP \cite{krieger}.

The application of the OEP methodology to the Fock term (\ref{fock}), 
either with or without the KLI approximation, is also known as the 
exact-exchange method (EXX). 
The OEP-EXX equations have been solved for atoms \cite{krieger,engelopm,grabo}
and solids \cite{kotani,exx2}, with very encouraging results.
Other orbital-dependent functionals that have been treated within the OEP
scheme are the PZ self-interaction correction \cite{sickli1,sickli2} and the 
Colle-Salvetti functional \cite{grabo}. A detailed review 
of the OEP and its KLI approximation is Ref.~\cite{oepreview}. 

The high accuracy attained by complex orbital functionals implemented via 
the OEP, and the fact that it is easier to devise orbital functionals than 
explicit density functionals, makes the OEP concept attractive, but the 
computational cost of solving the OEP integral equation is a major drawback. 
However, this computational cost is significantly reduced by the KLI 
approximation \cite{krieger} and other recently proposed 
simplifications \cite{kuemmelperdew,wuyang,elp}. In the context of the EXX 
method (i.e., using the Fock exchange term as orbital functional) the OEP is 
a viable way to proceed. For more complex orbital functionals, additional 
simplifications may be necessary \cite{krieger,kuemmelperdew,wuyang,elp}. 

A further reduction of computational complexity is achieved by not evaluating 
the orbital functional self-consistently, via Eq.~(\ref{oepinteg}), but only 
once, using the orbitals and densities of a converged self-consistent LDA or 
GGA calculation. This 
`post-GGA' or `post-LDA' strategy completely avoids the OEP and has been 
used both for hybrid functionals and Meta-GGAs \cite{b3,b10,metagga,bmeta}. 
A drawback of post methods is that they provide only approximations to
the selfconsistent total energies, not to eigenvalues, effective potentials,
orbitals or densities.

In the case of hybrid functionals, still another mode of implementation has 
become popular. This alternative, which also avoids solution of 
Eq.~(\ref{oepinteg}), is to calculate the derivative of the hybrid functional
with respect to the single-particle orbitals, and not with respect to the 
density as in (\ref{oepinteg}). The resulting single-particle equation 
is of Hartree-Fock form, with a nonlocal potential, and with a weight factor 
in front of the Fock term.  Strictly speaking, the orbital derivative is not 
what the HK theorem demands, but rather a Hartree-Fock like procedure, but
in practice it is a convenient and successful approach. This scheme, in which 
self-consistency is obtained with respect to the single-particle orbitals, 
can be considered an evolution of the Hartree-Fock Kohn-Sham method 
\cite{parryang}, and is how hybrids are commonly implemented. Recently, it 
has also been used for Meta-GGAs \cite{metaggatests}. For occupied orbitals, 
results obtained from orbital selfconsistency differ little from those 
obtained from the OEP.

Apart from orbital functionals, which are implicit nonlocal density 
functionals because the orbitals depend on the density in a nonlocal way, 
there is also a class of explicit nonlocal density functionals. Such
nonlocal density functionals take into account, at any point ${\bf r}$, not
only the density at that point, $n\r$, and its derivatives, $\nabla n\r$ etc.,
but also the behaviour of the density at different points ${\bf r}' \neq
{\bf r}$, by means of integration over physically relevant regions of
space. A typical example is
\be
E_{xc}^{ADA}[n] = \int d^3r \, n\r \epsilon_{xc}^{hom}(\bar{n}\r),
\ee
where $\epsilon_{xc}^{hom}$ is the per-particle $xc$ energy of the
homogeneous electron liquid (see footnote 47).
In the LDA one would have $\bar{n}\r \equiv n\r$, but in the average-density
approximation (ADA) one takes \cite{adawda}
\be
\bar{n}\r = \int d^3r' \, n\rp w[n](|{\bf r}-{\bf r}'|),
\label{nbar}
\ee
where $w[n](|{\bf r}-{\bf r}'|)$ is a weight function that samples the 
density not only semilocally, as do the GGAs, but over a volume determined 
by the range of $w$. Conceptually similar to the ADA is the weighted-density 
approximation (WDA) \cite{adawda}. In terms of the pair-correlation function
(see Secs. \ref{densmat} and \ref{corrint}) the LDA, ADA and WDA functionals
can be written as
\bea
E_{xc}^{LDA}[n]={e^2 \over 2}\int d^3r \int d^3r'\, 
\frac{n\r n\r}{|{\bf r}-{\bf r'}|}\left(\bar{g}_{hom}[n\r]({\bf r}-{\bf r'})-1 
\right)\\
E_{xc}^{ADA}[n]={e^2 \over 2}\int d^3r \int d^3r'\,
\frac{n\r \bar{n}\r}{|{\bf r}-{\bf r'}|}\left(\bar{g}_{hom}[\bar{n}\r]({\bf r}-{\bf r'})-1 \right)\\
E_{xc}^{WDA}[n]={e^2 \over 2}\int d^3r \int d^3r'\,
\frac{n\r n\rp}{|{\bf r}-{\bf r'}|}\left(\bar{g}_{hom}[\bar{n}\r]({\bf r}-{\bf r'})-1 \right),
\eea
where in each case $\bar{g}_{hom}({\bf r}-{\bf r'})$ is the pair-correlation
function of the homogeneous electron liquid, averaged over the 
coupling constant $e^2$ \cite{dftbook,parryang}.

The dependence of these functionals on $\bar{n}\r$, the integral over 
$n\r$, instead of on derivatives, as in the GGAs, is the reason why such
functionals are called nonlocal. In practice, this integral turns the
functionals computationally expensive, and in spite of their great promise
they are much less used than GGAs. However, recent comparisons of ADA and WDA 
with LDA and GGAs for low-dimensional systems \cite{adatest,wdatest} and for 
bulk silicon \cite{wdaadatest} show that nonlocal integral-dependent density 
functionals can outperform local and semilocal approximations.

\section{Extensions of DFT: New frontiers and old problems}
\label{extensions}
\hspace{10mm}

Up to this point we have discussed DFT in terms of the charge (or particle)
density $n\r$ 
as fundamental variable. In order to reproduce the correct charge density 
of the interacting system in the noninteracting (Kohn-Sham) system, one must
apply to the latter the effective KS potential $v_s=v+v_H+v_{xc}$, in which
the last two terms simulate the effect of the electron-electron interaction
on the charge density. This form of DFT, which is the one proposed originally 
\cite{hk}, could also be called `charge-only' DFT. It is not the most widely 
used DFT in practical applications. Much more common is a formulation that
employs one density for each spin, $n_\ua\r$ and $n_\da\r$, i.e, works with
two fundamental variables. In order to reproduce both of these in the
noninteracting system one must now apply two effective potentials,
$v_{s,\ua}\r$ and $v_{s,\da}\r$.\footnote{More generally, one requires one
effective potential for each density-like quantity to be reproduced in the
KS system. Such potentials and corresponding densities are called conjugate
variables.} This formulation of DFT is known as spin-DFT (SDFT) \cite{vbh,gl}.
Its fundamental variables $n_\ua\r$ and $n_\da\r$ can be used to calculate
the charge density $n\r$ and the spin-magnetization density $m\r$ from
\bea
n\r=n_\ua\r+n_\da\r \\
m\r=\mu_0(n_\ua\r-n_\da\r),
\eea
where $\mu_0=q\hbar/2mc$ is the Bohr magneton. More generally, the 
Hohenberg-Kohn theorem of SDFT states that in the presence of a magnetic 
field $B\r$ that couples only to the electron spin [via the familiar
Zeeman term $\int d^3r\,m\r B\r$] the ground-state wave function and all 
ground-state observables are unique functionals of $n$ and $m$ or,
equivalently, of $n_\ua$ and $n_\da$.\footnote{In the particular case $B=0$ 
the SDFT HK theorem still holds and continues to be useful, e.g., for systems 
with spontaneous polarization. In principle one could also use `charge-only'
DFT to study such systems, but then $n_\ua\r$ and $n_\da\r$ become functionals 
of $n\r$ and nobody knows how to determine these functionals.}
Almost the entire further development of the HK theorem and the KS 
equations can be immediately rephrased for SDFT, just by attaching a 
suitable spin index to the densities. For this reason we could afford the
luxury of exclusively discussing `charge-only' DFT in the preceding sections, 
without missing any essential aspects of SDFT.

There are, however, some exceptions to this simple rule. One is the fourth
statement of the HK theorem, as discussed in Sec.~\ref{hktheorem}. Another
is the construction of functionals. For the exchange energy it is known,
e.g., that \cite{oliver}
\be
E_x^{SDFT}[n_\ua,n_\da] = 
\frac{1}{2}\left( E_x^{DFT}[2n_\ua] + E_x^{DFT}[2n_\da] \right).
\label{spinscaling}
\ee
In analogy to the coordinate scaling of Eqs.~(\ref{cooscaling1}) -
(\ref{cooscaling3}), this property is often called `spin-scaling', and 
it can be used to construct an SDFT exchange functional from a given DFT 
exchange functional. In the context of the LSDA, von Barth and Hedin 
\cite{vbh} wrote the exchange functional in terms of an interpolation 
between the unpolarized and fully polarized electron gas which by
construction satisfies Eq.~(\ref{spinscaling}). Alternative interpolation
procedures can be found in Ref.~\cite{vwn}. GGA exchange functionals
also satisfy Eq.~(\ref{spinscaling}) by construction. For the correlation
energy no scaling relation of the type (\ref{spinscaling}) holds, so that
in practice correlation functionals are either directly constructed in 
terms of the spin densities or written by using, without formal justification,
the same interpolation already used in the exchange functional. In the case
of the LSDA this latter procedure was introduced in Ref.~\cite{vbh}, and 
further analysed and improved in Ref.~\cite{vwn}.

The Kohn-Sham equations of SDFT are
\be
\left[ -\frac{\hbar^2 \nabla^2}{2m} + v_{s\s}\r \right] \phi_{i\s}\r 
= \eps_{i\s} \phi_{i\s}\r,
\label{sdftkseq}
\ee
where $v_{s\s}\r = v_\s\r + v_H\r + v_{xc,\s}\r$. In a nonrelativistic
calculation the Hartree term does not depend on the spin label,\footnote{Spin-spin dipolar interactions are a relativistic effect of order $(1/c)^2$, as
are current-current interactions.} but in the
presence of an externally applied magnetic field $v_\s\r = v\r - \s \mu_0 B$
(where $\s =\pm 1$). Finally, 
\be
v_{xc,\s}\r = \frac{\delta E_{xc}^{SDFT}[n_\ua,n_\da]}{\delta n_\s\r}. 
\ee
In the presence of an internal magnetic field $B_{xc}$ (i.e., in 
spin-polarized systems) $v_{xc,\da} - v_{xc,\ua} =\mu_0 B_{xc}$.
This field is the origin of, e.g., ferromagnetism in transition metals.
References to recent work {\em with} SDFT include almost all practical DFT
calculation: SDFT is by far the most widely used form of DFT.\footnote{SDFT has become synonymous with DFT to such an extent that often no distinction is made
between the two, i.e., a calculation referred to as a DFT one is most of the
time really an SDFT one.} Some recent work {\em on} SDFT is described
in Ref.~\cite{sanibel}.
A more detailed discussion of SDFT can be found in 
Refs.~\cite{dftbook,parryang,gl}, and a particularly clear exposition of
the construction of $xc$ functionals for SDFT is the contribution of
Kurth and Perdew in Refs.~\cite{joulbert,primer}.

If the direction of the spins is not uniform in space\footnote{Such
`noncollinear magnetism' appears, e.g., as canted or helical spin 
configurations in rare-earth compounds, or as domain walls in
ferromagnets.} one requires a formulation of SDFT in which the spin 
magnetization is not a scalar, as above, but a three-component vector
${\bf m}\r$. Different proposals for extending SDFT to this situation
are available \cite{sandratskii,nordstroem,sdwprb}. One mechanism that 
can give rise to noncollinear magnetism is spin-orbit coupling. This is
another relativistic effect \cite{strange}, and as such it is 
not consistently treated in either DFT or SDFT. A generalization of DFT
that does account for spin-orbit coupling and other relativistic effects
is relativistic DFT (RDFT) \cite{rajagopal,macdonald}. Here the fundamental
variable is the relativistic four-component current $j^\mu$. RDFT requires 
a more 
drastic reformulation of DFT than does SDFT. In particular, the KS equation
of RDFT is now of the form of the single-particle Dirac equation, instead of
the Schr\"odinger equation. There are also many subtle questions involving
renormalizability and the use of the variational principle in the presence
of negative energy states. For details on these problems and their eventual
solution the reader is referred to the chapters by Engel {\em et al.} in 
Refs.~\cite{nalewajski} and \cite{asi95}, and to the book by Eschrig 
\cite{eschrig}. 
A didactical exposition of RDFT, together with representative applications
in atomic and condensed-matter physics, can be found in the book by Strange 
\cite{strange}, and a recent numerical implementation is presented in 
Ref.~\cite{engelrel}.

To study the magnetic properties of matter one would often like to be able to
obtain information on the currents in the system and their coupling to 
possible external magnetic fields. Important classes of experiments for which
this information is relevant are nuclear magnetic resonance and the
quantum Hall effects. SDFT does not provide explicit information on the
currents. RDFT in principle does, but standard implementations of it are 
formulated in a spin-only version, which prohibits extraction of information 
on the currents. Furthermore, the formalism of RDFT is considerably more
complicated than that of SDFT. 
In this situation the formulation of nonrelativistic current-DFT (CDFT),
accomplished by Vignale and Rasolt \cite{vr1,vr2}, was a major step forward.
CDFT is formulated explicitly in terms of the (spin) density and the
nonrelativistic paramagnetic current density vector ${\bf j}_p\r$.
Some recent applications of CDFT are Refs.~\cite{ferconi,ebert,handy,jpe}.
E.~K.~U.~Gross and the author have shown that the existence of {\it spin 
currents} implies the existence of a link between the $xc$ functionals of SDFT 
and those of CDFT \cite{cdftlett}. Conceptually, this link is similar 
to the one of Eq.~(\ref{spinscaling}) between functionals of DFT and SDFT,
but the details are quite different. Some approximations for $xc$ functionals 
of CDFT are discussed in Refs.~\cite{cdftlett,skudlarski,pcdft}.

In addition to SDFT, RDFT and CDFT, there exist many other generalizations
of DFT that were designed for one or other special purpose. As examples
we mention superconductivity \cite{ogk1,ogk2,dftscprlbbalazs,dftscldaprl}
and spin-density waves \cite{sdwprb,sdwlett}, but there are many more 
[5-19]. For reasons of space we cannot discuss these
extensions here. Instead, let us take a brief look at a problem that requires
more radical departures from the framework of conventional DFT: excited
states. DFT is formulated in terms of ground-state densities, and it is
not immediately obvious how one could extract information on excited states
from them (although at least in the case of `charge-only' DFT the fourth
substatement of the HK theorem guarantees that this must be possible).

Apart from the {\it ad hoc} identification of the KS eigenvalues with true 
excitation energies, there exists a considerable variety of more sound 
approaches to
excited states in DFT that have met with some success. The early suggestion
of Gunnarsson and Lundqvist \cite{gl} to use a symmetry-dependent $xc$
functional to calculate the lowest-energy excited state of each symmetry
class has been implemented approximately by von~Barth \cite{vonbarth}, but
suffers from lack of knowledge on the symmetry dependence of the functional.
More recent work on this dependence is Ref.~\cite{goerlingsym}.
An alternative approach to excited states, not restricted to the lowest energy 
state of a given symmetry, is ensemble DFT, developed by Theophilou 
\cite{theophilou} and further elaborated by Oliveira, Gross and Kohn 
\cite{ensembles}.
In this formalism the functional depends on the particular choice for the 
ensemble, and a simple approximation for this dependence is available
\cite{ensembles}. Some applications of this method have been worked out by 
Nagy \cite{nagy}. 

Other DFT approaches to excited states can be found in 
Refs.~\cite{goerlingex}, \cite{nagyex}, \cite{gencoopaper} and 
\cite{perdewlevy}, but the most widely used method today is time-dependent 
DFT (TD-DFT). The time-dependent 
generalization of the HK theorem, the Runge-Gross theorem, cannot be
proven along the lines of the original HK theorem, but requires a different
approach \cite{rungegross,robert}. For recent reviews of TD-DFT see
Ref.~\cite{tdreview}. Excited states have first been extracted from TD-DFT 
in Refs.~\cite{ex1,ex2}. This approach is now implemented in standard 
quantum-chemical DFT program packages \cite{baerends,scuseria} and is 
increasingly applied also in solid-state physics \cite{rubio}.
Another important application of TD-DFT is to systems in external
time-dependent fields, such as atoms in strong laser fields
\cite{laser1,laser2}. First steps towards studying dynamical magnetic
phenomena with TD-SDFT have been taken very recently \cite{spindynlett}. 

All these extensions of DFT to time-dependent, magnetic, relativistic 
and a multitude of other situations involve more complicated Hamiltonians
than the basic {\it ab initio} many-electron Hamiltonian defined by 
Eqs.~(\ref{mbse}) to (\ref{expotl}). Instead of attempting to achieve a 
more complete description of the many-body system under study by adding
additional terms to the Hamiltonian, it is often advantageous to employ
the opposite strategy, and reduce the complexity of the {\it ab initio} 
Hamiltonian by replacing it by simpler models, which focalize on specific 
aspects of the full many-body problem. Density-functional theory can be 
applied to such model Hamiltonians, too, once a suitable density-like
quantity has been identified as basic variable.
Following pioneering work by Gunnarsson and Sch\"onhammer \cite{gshub},
LDA-type approximations have, e.g., recently been formulated and exploited
for the Hubbard \cite{balda}, the delta-interaction \cite{magyar}
and the Heisenberg \cite{hemo} models. Common aspects and potential
uses of DFT for model Hamiltonians are described in \cite{hechap}.

Still another way of using DFT, which does not depend directly on
approximate solution of Kohn-Sham equations, is the quantification and
clarification of traditional chemical concepts, such as electronegativity
\cite{parryang}, hardness, softness, Fukui functions, and other reactivity
indices \cite{parryang,chermette}, or aromaticity \cite{geerlings}.
The true potential of DFT for this kind of investigation is only beginning
to be explored, but holds much promise.

All extensions of DFT face the same formal questions (e.g., simultaneous
interacting and noninteracting $v$-represen\-ta\-bility of the densities, 
nonuniqueness of the KS potentials, meaning of the KS eigenvalues) and 
practical problems (e.g., how to efficiently solve the KS equations, how to 
construct accurate approximations to $E_{xc}$, how to treat systems with very 
strong correlations) as do the more widely used formulations 
`charge-only' DFT and SDFT.
These questions and problems, however, have never stopped DFT from
advancing, and at present DFT emerges as the method of choice for solving 
a wide variety of quantum mechanical problems in chemistry and physics --- 
and in many situations, such as large and inhomogeneous systems, it is the
only applicable first-principles method at all.

The future of DFT is bright \cite{rmp1,kohnbeckeparr,mattsson} 
 --- but to be able to contribute to it, the reader
must now leave the present superficial overview behind, and turn to the
more advanced treatments available in the literature [5-19].\\

{\bf Acknowledgments}\,
The author has learned density-functional theory from E.~K.~U. Gross, 
and then practiced it in collaborations with B.~L.~Gy\"orffy, L.~N.~Oliveira, 
and G.~Vignale. These scholars are in no way responsible for the content of 
this work, but the author's intellectual debt to them is enormous.
Useful comments by J.~Quintanilla, H.~J.~P.~Freire, T.~Marcasso, E.~Orestes, 
N.~A.~Lima, N.~Argaman, V.~L.~L\'{\i}bero, V.~V.~Fran\c{c}a, M.~Odashima, 
J.~M.~Morbec, A.~P.~F\'avaro, A.~J.~R.~da Silva and L.~N.~Oliveira on earlier 
versions of this manuscript are gratefully acknowledged. This work was 
supported financially by FAPESP and CNPq.

%]})
\end{document}